\documentclass[preprint2]{proto}
\usepackage{times}

\def\ni{\noindent}
\def\s{{\rm\,s}}

\def\cm{{\rm\,cm}}
\def\m{{\rm\,m}}
\def\km{{\rm\,km}}

\def\gm{{\rm\,g}}

\def\pc{{\rm\,pc}}

\def\AU{{\rm\, AU}}

\def\yr{{\rm\,yr}}
\def\Myr{{\rm\,Myr}}
\def\Gyr{{\rm\,Gyr}}

\newcommand{\refs}{\par\noindent\hangindent=1pc\hangafter=1}

\begin{document}

\title{\textbf{\LARGE A Brief History of Trans-Neptunian Space}}

\author {\textbf{\large Eugene Chiang, Yoram Lithwick, and Ruth Murray-Clay}}
\affil{\small\em University of California at Berkeley}
\author {\textbf{\large Marc Buie and Will Grundy}}
\affil{\small\em Lowell Observatory}
\author{ \textbf{\large Matthew Holman}}
\affil{\small\em Harvard-Smithsonian Center for Astrophysics}

\begin{abstract}
\begin{list}{ } {\rightmargin 0.4in}
\baselineskip = 11pt
\parindent=1pc
{\small The Edgeworth-Kuiper belt
encodes the dynamical history of the outer solar system. Kuiper belt objects
(KBOs)
bear witness to coagulation physics, the evolution of planetary orbits,
and external perturbations from the solar neighborhood.
We critically review the present-day belt's observed properties
and the theories designed to explain them.
Theories are organized according to a possible time-line
of events. In chronological order, epochs described include
(1) coagulation of KBOs in a dynamically cold disk,
(2) formation of binary KBOs by fragmentary collisions and gravitational
captures,
(3) stirring of KBOs by Neptune-mass planets (``oligarchs''),
(4) eviction of excess oligarchs,
(5) continued stirring of KBOs
    by remaining planets whose orbits circularize by dynamical friction,
(6) planetary migration and capture of Resonant KBOs,
(7) creation of the inner Oort cloud by passing stars in an open stellar
cluster, and
(8) collisional comminution of the smallest KBOs.
Recent work underscores how small, collisional, primordial planetesimals
having low velocity dispersion permit the rapid assembly of $\sim$5
Neptune-mass
oligarchs at distances of 15--25 AU. We explore the consequences of such
a picture. We propose that Neptune-mass planets
whose orbits cross into the Kuiper belt for up to $\sim$20 Myr
help generate the high-perihelion members
of the hot Classical disk and Scattered belt.
By contrast, raising perihelia by sweeping secular resonances during
Neptune's migration might fill these reservoirs too inefficiently when account
is made of how
little primordial mass might reside in bodies having sizes on the order of 100
km. These and other frontier issues in trans-Neptunian space are discussed
quantitatively.
 \\~\\~\\~\\}

\end{list}
\end{abstract}  

\noindent
\centerline{\textbf{1. INTRODUCTION}}

\bigskip
The discovery by {\it Jewitt and Luu} (1993) of what many now regard as the
third
Kuiper belt object opened a new frontier in planetary astrophysics: the direct
study of
trans-Neptunian space, that great expanse extending beyond the orbit of the
last known giant planet
in our solar system. This space is strewn with icy, rocky bodies having
diameters ranging up to
a few thousand km and occupying orbits of a formerly unimagined variety.

Kuiper belt objects (KBOs) afford insight into processes that form and shape
planetary systems. In contrast to main belt asteroids, the largest
KBOs today have lifetimes against collisional disruption that well exceed the
age
of the universe. Therefore their size spectrum may
preserve a record, unweathered by erosive collisions, of the
process by which planetesimals and planets coagulated.
At the same time, KBOs can be considered test particles whose trajectories have
been
evolving for billions of years in a time-dependent gravitational potential.
They provide intimate testimony of how the giant planets---and perhaps even
planets
that once resided within our system but have since been ejected---had their
orbits sculpted.
The richness of structure revealed by studies of our homegrown debris disk
is unmatched by more distant, extra-solar analogues.

Section 2 summarizes observed properties of the Kuiper belt.
Some of the data and analyses concerning orbital elements
and spectral properties of KBOs are new and have not been published
elsewhere. Section 3 is devoted to theoretical interpretation.
Topics are treated in order of a possible chronology of events in the outer
solar system.
Parts of the story that remain missing or that are contradictory
are identified. Section 4 recapitulates
a few of the bigger puzzles.

Our review is packed with simple and hopefully illuminating
order-of-magnitude calculations that readers are encouraged to reproduce or
challenge.
Some of these confirm claims made in the literature
that would otherwise find no support apart from numerical simulations.
Many estimates are new, concerning all the ways in which Neptune-sized
planets might have dynamically excited the Kuiper belt.
While we outline many derivations,
space limitations prevent us from spelling out all details.
For additional guidance, see the pedagogical review
of planet formation by {\it Goldreich et al.}~(2004a, hereafter {\it G}04),
from which our work draws liberally.

\bigskip
\centerline{\textbf{2. THE KUIPER BELT OBSERVED TODAY}}
\bigskip

\centerline{\textbf{2.1. Dynamical Classes}}
\bigskip

Outer solar system objects divide into dynamical classes
based on how their trajectories evolve. Fig.~1 displays orbital elements,
time-averaged
over 10 Myr in a numerical orbit integration that accounts for the masses
of the four giant planets,
of 529 objects. Dynamical~classifications of these objects are secure according
to criteria
developed by the Deep Ecliptic Survey (DES) team ({\it Elliot~et~al.},~2005,
hereafter {\it E}05).
Fig.~2 provides a close-up view of a portion of the Kuiper Belt.
We distinguish 4 classes:

1. {\it Resonant KBOs} (122/529)
exhibit one or more mean-motion commensurabilities
with Neptune, as judged by steady libration of the appropriate
resonance angle(s) ({\it Chiang et al.}, 2003, hereafter {\it C}03).
Resonances most heavily populated include 
the exterior 3:2 (Plutino), 2:1, and 5:2; see Table 1.
Of special interest is the first discovered Neptune Trojan (1:1).
All Resonant KBOs (except the Trojan)
are found to occupy $e$-type resonances; the ability of an $e$-type resonance
to retain a KBO tends to increase with the KBO's eccentricity $e$
(e.g., {\it Murray and Dermott}, 1999). Unless otherwise
stated, orbital elements are heliocentric and referred to the invariable
plane.
Several (9/122) also inhabit inclination-type ($i^2$) or mixed-type ($ei^2$)
resonances.
None inhabits an $e_{\rm N}$-type resonance whose stability
depends on the (small) eccentricity of Neptune. The latter observation
is consistent with numerical experiments that suggest $e_{\rm N}$-type
resonances are rendered unstable by adjacent $e$-type resonances.

\begin{table}[h]
\begin{center}
Table~1.---Observed Populations of Neptune Resonances

(securely identified by the DES team as of 8 Oct 2005)
\end{center}

\begin{tabular}{cc|cc|cc|cc|cc} \tableline \tableline
\multicolumn{2}{c}{Order 0} &
\multicolumn{2}{|c}{Order 1} &
\multicolumn{2}{|c}{Order 2} &
\multicolumn{2}{|c}{Order 3} &
\multicolumn{2}{|c}{Order 4} \\
m:n & \# & m:n & \# & m:n & \# & m:n & \# & m:n & \# \\ \tableline
1:1 & 1       & 5:4     & 4   & 5:3     & 9   & 7:4     & 8   & 9:5     & 2
\\
    &         & 4:3     & 3   & 3:1     & 1   & 5:2     & 10  & 7:3     & 1
\\
    &         & 3:2     & 72  &         &     &         &     &         &
\\
    &         & 2:1     & 11  &         &     &         &     &         &
\\ \tableline
\end{tabular}
\end{table}


2. {\it Centaurs} (55/529) are non-Resonant objects whose perihelia
penetrate inside the orbit of Neptune. Most Centaurs cross the Hill sphere
of a planet within 10 Myr. Centaurs are likely descendants of the other 3
classes, recently dislodged from
the Kuiper belt by planetary perturbations
({\it Holman and Wisdom}, 1993, hereafter {\it HW}93; {\it Tiscareno and
Malhotra}, 2003). They
will not be discussed further.

\begin{figure*}[h!]
\vspace{-4in}
 \epsscale{1.7}
 \plotone{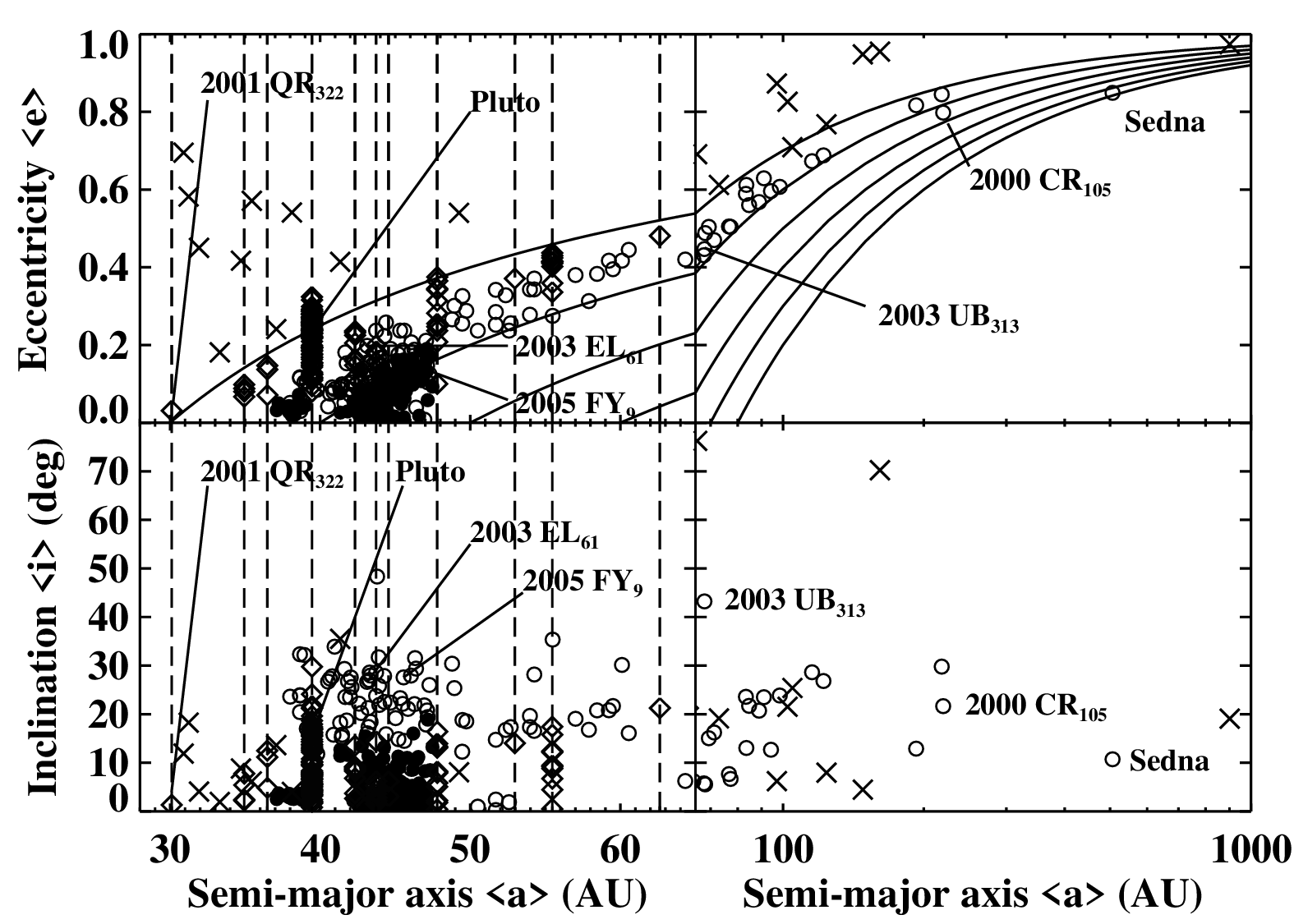}
 \caption{\small Orbital elements, time-averaged over 10 Myr, of 529 securely
classified outer solar system objects
as of 8 Oct 2005. Symbols represent dynamical classes: Centaurs ($\times$),
Resonant KBOs ($\Diamond$), Classical KBOs ($\bullet$), and
Scattered KBOs ($\circ$).
Dashed vertical lines indicate occupied mean-motion resonances;
in order of increasing heliocentric distance, these include
the 1:1, 5:4, 4:3, 3:2, 5:3, 7:4, 9:5, 2:1, 7:3, 5:2, and 3:1
(see Table 1).
Solid curves trace loci of constant perihelion $q = a(1-e)$. 
Especially large (2003 UB$_{313}$, Pluto, 2003 EL$_{61}$, 2005 FY$_{9}$; {\it
Brown et al.}, 2005ab)
and dynamically unusual KBOs are labelled (2001 QR$_{322}$ [Trojan; {\it Chiang
et al.}, 2003; {\it Chiang and Lithwick}, 2005],
2000 CR$_{105}$ [high $q$; {\it Millis et al.}, 2002; {\it Gladman et al.},
2002], Sedna [high $q$; {\it Brown et al.}, 2004]). For a zoomed-in view, see
Figure 2.
}
 \end{figure*}

\begin{figure}[ht]
\epsscale{1}
\plotone{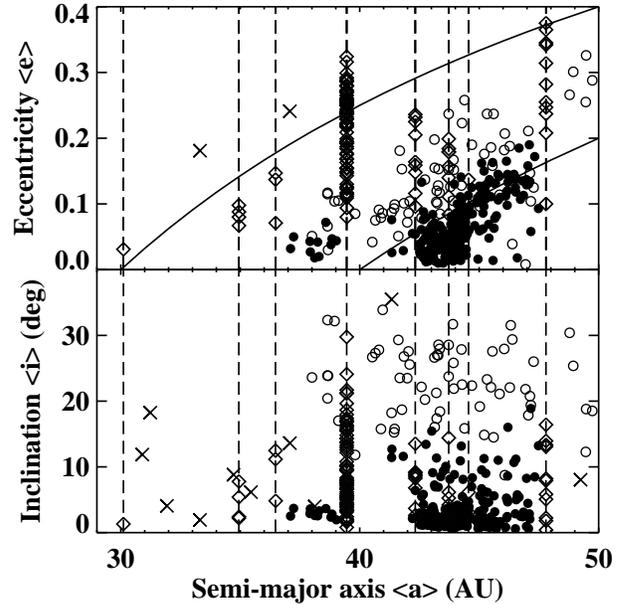}
\caption{\small Same as Fig.~1, zoomed in.}
\end{figure}

3. {\it Classical KBOs} (246/529) are non-Resonant, non-planet-crossing
objects
whose time-averaged $\langle e \rangle \leq 0.2$
and whose time-averaged Tisserand parameters

\begin{equation}
\langle T \rangle = \langle (a_{\rm N}/a) + 2 \sqrt{ (a/a_{\rm N}) (1-e^2)}
\cos \Delta i \rangle \,
\end{equation}

\ni exceed 3. Here $\Delta i$ is the mutual inclination between the orbit
planes
of Neptune and the KBO, $a$ is the semi-major axis of the KBO, and $a_{\rm N}$
is the semi-major axis of Neptune.
In the circular, restricted, 3-body problem, test particles with $T > 3$ and $a
> a_{\rm N}$
cannot cross the orbit of the planet (i.e., their perihelia $q = a(1-e)$
remain greater than $a_{\rm N}$).
Thus, Classical KBOs can be argued
to have never undergone close encounters with Neptune in its current nearly
circular
orbit and to be relatively pristine dynamically.
Indeed, many Classical KBOs as identified by our scheme
have low inclinations $\langle i \rangle < 5^{\circ}$ (``cold Classicals''),
though some do not (``hot Classicals''). Our defining threshold for
$\langle e \rangle$ is arbitrary; like our threshold for $\langle T \rangle$,
it is imposed to suggest---perhaps incorrectly---which
KBOs might have formed and evolved {\it in situ}.

Classical KBOs have spectral properties distinct from those of other
dynamical classes: Their colors
are more uniformly red (Fig.~3; see the chapter by {\it Cruikshank et al.}).
According to the Kolmogorov-Smirnov test ({\it Press et al.}, 1992; {\it
Peixinho et al.}, 2004), the
probabilities that Classical KBOs have $\bv$ colors and Boehnhardt-$S$
slopes ({\it Boehnhardt et al.}, 2001) drawn from distributions identical to
those for Resonant KBOs are $10^{-2}$ and $10^{-3}$, respectively.
When Classical KBOs are compared to Scattered KBOs (see below), the
corresponding probabilities are $10^{-6}$ and $10^{-4}$.  
An alternative interpretation is that low-$i$ KBOs
are redder than high-$i$ KBOs ({\it Trujillo and Brown}, 2002; {\it Peixinho et
al.}, 2004).
This last claim
is statistically significant when Classical, Scattered, and Resonant KBOs
are combined and analyzed as one set (Fig.~4). However, no correlation between
physical properties and $i$ (or any other measure of excitation)
has proven significant within any individual class.

%


\begin{figure}[ht]
\epsscale{1}
\plotone{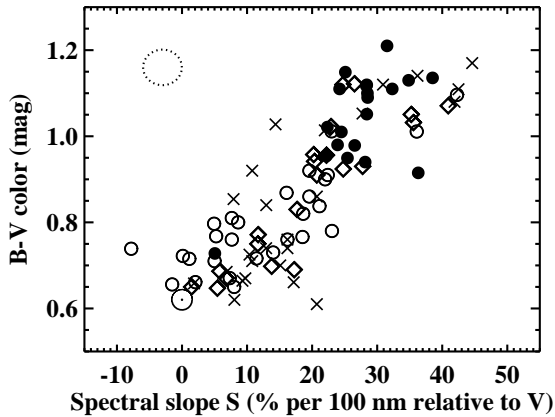}
\caption{\small Visual colors of KBOs and Centaurs calculated from published
photometry, with the average uncertainty indicated by
the upper left oval.  The spectral slope $S$ is calculated for
wavelengths in the range of Johnson $V$ through $I$.
Neutral (solar) colors are indicated by $\odot$.
Symbols for dynamical classes are the same as those for Fig.~1.
Classical KBOs constitute a distinct red cluster,
except for (35671), which also has a
small semi-major axis of 38 AU. Other classes are widely dispersed
in color.}
\end{figure}

Both the inner edge of the Classical belt at $a \approx 37 \AU$,
and the gap in the Classical belt at $a \approx 40$--42 AU and
$\langle i \rangle \lesssim 10^{\circ}$ (see Fig.~2), reflect
ongoing sculpting by the present-day planets.
The inner edge marks the distance out to which the planets have
eroded the Kuiper belt over the last few billion
years ({\it HW}93; {\it Duncan et al.}, 1995, hereafter {\it D}95).
The gap is carved by the $\nu_{18}$, $\nu_{17}$, and $\nu_8$ secular
resonances
({\it HW}93; {\it D}95; {\it Kne\v{z}evi\'{c} et al.}, 1991). At a secular
resonance
denoted by $\nu_j$, the orbital precession frequency of a test
particle---apsidal
if $j < 10$ and nodal if $j > 10$---matches one of the precession
eigenfrequencies
of the planets (see chapter 6 of {\it Murray and Dermott}, 1999).
For example, at low $i$, the $\nu_8$ resonance drives $e$
to Neptune-crossing values in $\sim$$10^6\yr$.
Particles having large $i$, however, can elude the $\nu_8$
({\it Kne\v{z}evi\'{c} et al.}, 1991).
Indeed, 18 KBOs of various classes and all having large $\langle i \rangle$
reside within the gap.

\begin{figure}[h]
\epsscale{1}
\plotone{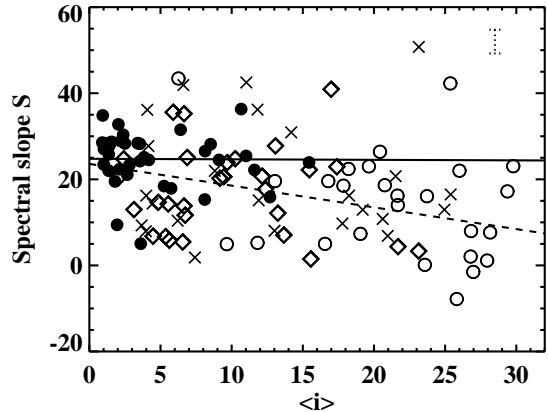}
\caption{\small
Spectral slope $S$ vs.~time-averaged inclination.
The typical uncertainty in $S$ is indicated by the dotted bar.
Classical KBOs evince no trend of color with $\langle i \rangle$.
The solid line is fitted to Classicals only;
statistical tests using Spearman's rank-order coefficient
and Kendall's tau ({\it Press et al.}, 1992) show that no significant
correlation
exists. When Classical, Resonant, and Scattered KBOs are combined,
$S$ and $\langle i \rangle$ correlate significantly (with a false
alarm probability of $10^{-5}$); the dashed
line is a fit to all three classes.
The two most neutral Classicals are (35671) and 1998 WV$_{24}$,
having semi-major axes of 38 and 39 AU, respectively.
}
\end{figure}

By contrast, the outer edge of the Classical belt at $a \approx 47 \AU$
is likely primordial. Numerous surveys
(e.g., {\it E}05; {\it Bernstein et al.}, 2004; and references therein)
carried out after an edge was initially suspected
({\it Jewitt et al.}, 1998) all failed to find a single object
moving on a low-$e$ orbit outside 47 AU.
The reality of the ``Kuiper Cliff'' is perhaps most convincingly demonstrated
by {\it Trujillo and Brown} (2001),
who simply plot the distribution of heliocentric discovery distances
of (mostly Classical) KBOs after correcting for the bias against
finding more distant, fainter objects.
This distribution peaks
at 44 AU and plummets to a value 10 times smaller at 50 AU.
The statistical significance of the Cliff hinges upon the fact
that the bias changes less dramatically---only by a factor
of 2.2--2.4 for reasonable parameterizations of the size
distribution---between 44 and 50 AU.
The possibility remains that predominantly small objects having
radii $R < 50 \km$ reside
beyond 47 AU,
or that the Cliff marks the inner edge of an annular gap
having radial width $\gtrsim 30\AU$ ({\it Trujillo and Brown}, 2001).

4. {\it Scattered KBOs} (106/529) comprise non-Classical, non-Resonant
objects whose perihelion distances $q$ remain outside the orbit of Neptune.
(The ``Scattered-Near'' and
``Scattered-Extended'' classes defined in {\it E}05---see also {\it Gladman et
al.}~(2002)---are
combined to simplify discussion. Also, while we do not formally introduce Oort
cloud
objects as a class, we make connections to that population throughout this
review.)
How were Scattered KBOs emplaced onto their highly elongated and inclined
orbits? Appealing to perturbations exerted by the giant planets in their
current orbital configuration
is feasible only for some Scattered objects.
A rule-of-thumb derived from numerical experiments
for the extent of the planets' collective reach is $q \lesssim 37 \AU$
({\it D}95; {\it Gladman et al.}, 2002).
Fig.~1 reveals that many Scattered objects possess
$q > 37 \AU$ and are therefore problematic. Outstanding examples include
2000 CR$_{105}$ ($q = 44 \AU$; {\it Millis et al.}, 2002; {\it Gladman et al.},
2002) and
(90377) Sedna ($q = 76 \AU$; {\it Brown et al.}, 2004).

These classifications are intended to sharpen analyses
and initiate discussion. The danger lies in allowing them
to unduly color our thinking about origins.
For example, though Sedna is classified above as a
Scattered KBO, the history of its orbit may be
distinct from those of other Scattered KBOs. We make
this distinction explicit below.

\bigskip
\centerline{\textbf{2.2. Sky Density and Mass}}
\bigskip

We provide estimates for the masses of the Kuiper belt
(comprising objects having $q \lesssim 60 \AU$ and $a > 30 \AU$; \S2.2.1);
the inner Oort Cloud (composed of Sedna-like objects; \S2.2.2);
and Neptune Trojans ($a \approx 30 \AU$; \S2.2.3).

\bigskip
\centerline{2.2.1. Main Kuiper Belt}
\bigskip

\begin{figure}[ht]
\epsscale{1}
\plotone{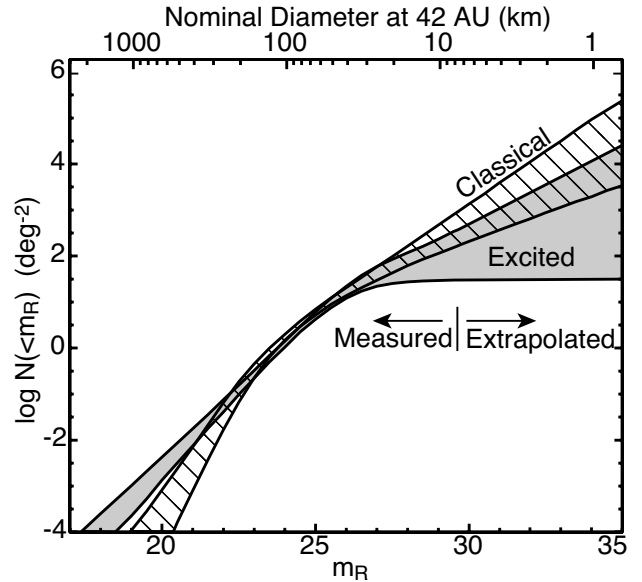}
\caption{\small Cumulative sky density vs.~apparent red magnitude for
CKBOs and Excited KBOs, from {\it B}04. If $N (< m_R) \propto 10^{\alpha
m_R}$,
then the size distribution index $\tilde{q} = 5 \alpha + 1$ (see \S2.2.1).
Envelopes enclose 95\% confidence intervals. The top abscissa
is modified from {\it B}04; here it assumes a visual albedo of 10\%.
Figure provided by Gary Bernstein.
\label{garyb}
}
\end{figure}

{\it Bernstein et al.}~(2004, hereafter {\it B}04) compile data from published
surveys
in addition to their own unprecedentedly deep Hubble Space Telescope (HST)
survey
to compute the cumulative sky density of KBOs versus apparent red magnitude
$m_R$ (``luminosity function''), shown in Fig.~5. Sky densities
are evaluated near the ecliptic plane.
Objects are divided into two groups: ``CKBOs'' (similar to our Classical
population) having heliocentric distances $38\AU < d < 55\AU$ and ecliptic
inclinations $i \leq 5^{\circ}$,
and ``Excited'' KBOs (similar to our combined Resonant and Scattered classes)
having $25 \AU < d < 60 \AU$ and $i > 5^{\circ}$.
Given these definitions, their analysis excludes
objects with ultra-high perihelia
such as Sedna.
With 96\% confidence, B04 determine that CKBOs and Excited KBOs
have different luminosity functions.
Moreover, neither function conforms
to a single power law from $m_R = 18$ to 29; instead, each is
well-fitted by a double power law that flattens towards fainter magnitudes.
The flattening
occurs near $m_R \approx 24$ for both groups.
To the extent that all objects in a group have the same albedo
and are currently located at the same
heliocentric distance, the luminosity function
is equivalent to the size distribution.
We define $\tilde{q}$ as the slope of the differential size distribution,
where
$dN \propto R^{-\tilde{q}} dR$ equals the number of objects
having radii between $R$ and $R+dR$. As judged from Fig.~\ref{garyb}, for
CKBOs,
$\tilde{q}$ flattens from $5.5$--$7.7$ (95\% confidence interval)
to $1.8$--$2.8$ as $R$ decreases. For Excited KBOs,
$\tilde{q}$ flattens from $4.0$--$4.6$ to $1.0$--$3.1$.
Most large objects are Excited
(see also Fig.~1).

By integrating the luminosity function over all magnitudes,
{\it B}04 estimate the total mass in CKBOs to be

\begin{eqnarray}
M_{\rm CKBO} \approx 0.005 \left( \frac{p}{0.10} \right)^{-3/2} \left(
\frac{d}{42 \AU} \right)^6 \times \nonumber \\
\left( \frac{\rho}{2 \gm \cm^{-3}} \right) \left( \frac{A}{360^{\circ} \times
6^{\circ}} \right) M_{\oplus} \,,
\end{eqnarray}

\ni where all CKBOs are assumed to have
the same albedo $p$,
heliocentric distance $d$,
and internal density $\rho$. The solid angle subtended by the belt of CKBOs is
$A$.
Given uncertainties in the
scaling variables---principally $p$ and $\rho$ (see the chapter by {\it
Cruikshank et al.}~for
recent estimates)---this mass is good to
within a factor of several. The mass is concentrated in objects
having radii $R \sim 50\km$, near the break in the luminosity
function.

The mass in Excited KBOs
cannot be as reliably calculated.
This is because the sample is heterogeneous---comprising
both Scattered and Resonant KBOs having a wide dispersion
in $d$---and because corrections need to be made for
the observational bias against finding objects near the aphelia of
their eccentric orbits. The latter
bias can be crudely quantified as $(Q/q)^{3/2}$:
the ratio of time an object spends near its aphelion distance
$Q$ (where it is undetectable) versus its perihelion distance $q$ (where it is
usually discovered).
An order-of-magnitude estimate that accounts
for how much larger $A$, $d^6$, and
$(Q/q)^{3/2}$ are for Excited KBOs than for CKBOs
suggests that the former population might weigh
$\sim$10 times as much as the latter, or $\sim$$0.05 M_{\oplus}$.
This estimate assumes that $\tilde{q}$ for Excited KBOs
is such that most of the mass is concentrated near $m_R \approx 24$,
as is the case for CKBOs.
If $\tilde{q}$ for the largest Excited KBOs is as small
as 4, then our mass estimate increases by
a logarithm to $\sim$$0.15 M_{\oplus}$.

\bigskip
\centerline{2.2.2. Inner Oort Cloud (Sedna-Like Objects)}
\bigskip

What about objects with unusually high perihelia such as Sedna,
whose mass is
$M_{\rm S} \sim 6 \times 10^{-4} (R/750 \km)^3
M_{\oplus}$?
The Caltech Palomar Survey searched $f \sim 1/5$
of the celestial sphere to discover one such object ({\it Brown et al.},
2004).
By assuming Sedna-like objects are distributed isotropically
(not a forgone conclusion; see \S3.7),
we derive an upper limit
to their total mass of $M_{\rm S} (Q/q)^{3/2} f^{-1} \sim 0.1 M_{\oplus}$.
If all objects on Sedna-like orbits obey
a size spectrum resembling that of Excited KBOs ({\it B}04), then we
revise the upper limit to $\sim$$0.3 M_{\oplus}$.
The latter value is 20 times smaller than the estimate
by {\it Brown et al.}~(2004); the difference arises from our use
of a more realistic size distribution.

\bigskip
\centerline{2.2.3. Neptune Trojans}
\bigskip

The first Neptune Trojan, 2001 QR$_{322}$ (hereafter ``QR''), was
discovered by the DES ({\it C}03).
The distribution of DES search fields on the sky, coupled with theoretical
maps of the sky density of Neptune Trojans ({\it Nesvorn\'{y} and Dones},
2002),
indicate that $N \sim 10$--30 objects resembling QR librate on tadpole
orbits about Neptune's forward Lagrange (L4) point ({\it C}03). Presumably
a similar population exists at L5.
An assumed albedo of 12--4\% yields a radius for QR
of $R \sim 65$--115 km.
Spreading the inferred population of QR-like objects
over the area swept out by tadpole orbits gives a surface mass
density in a single
Neptune Trojan cloud that approaches that of the main Kuiper belt to within
factors of a few ({\it Chiang and Lithwick}, 2005). Large Neptune Trojans
are at least comparable in number to large Jupiter Trojans and may outnumber
them by a factor of $\sim$10 ({\it C}03).

\bigskip
\centerline{\textbf{2.3. Binarity}}
\bigskip

{\it Veillet et al.}~(2002) optically resolved the first binary (1998
WW$_{31}$)
among KBOs having sizes $R \sim 100 \km$.
Over twenty binaries having components in this size range are now resolved.
Components typically have comparable brightnesses and are separated (in
projection)
by 300--10$^5$ km (e.g., {\it Stephens and Noll}, 2006).
These properties reflect observational biases against
resolving binaries that are separated by $\lesssim$ 0.1$\arcsec$
and that contain faint secondaries.

Despite these selection biases, {\it Stephens and Noll} (2006) resolved
as many as 9 out of 81 KBOs ($\sim$10\%) with HST.
They further report that the incidence of binarity appears 4 times higher
in the Classical disk than in other dynamical populations.
It is surprising that so many binaries exist with components widely separated
and comparably sized. A typical binary in the asteroid belt, by contrast,
comprises dissimilar masses separated by distances only slightly
larger than the primary's radius. Another peculiarity of
KBO binaries is that components orbit one other with
eccentricities of order unity. In addition, binary orbits are inclined
relative
to their heliocentric orbits at seemingly random angles.
See {\it Noll} (2003) for more quantitative details.

Although close binaries cannot be resolved, their components can
eclipse each other. {\it Sheppard and Jewitt} (2004) highlight a system whose
light curve varies with large amplitude and little
variation in color---suggesting that it is a near-contact binary.
They infer that at least 10\% of KBOs are members of similarly close binaries.

Among the four largest KBOs having $R \approx 1000 \km$---2003 UB$_{313}$,
Pluto,
2005 FY$_9$, and 2003 EL$_{61}$---three are known to possess satellites.
Secondaries for 2003 EL$_{61}$
and 2003 UB$_{313}$ are 5\% and 2\% as bright as their primaries, and
separated
by 49500 and 36000 km, respectively ({\it Brown et al.}, 2005b).
In addition to harboring Charon ({\it Christy and Harrington}, 1978), Pluto
possesses two, more distant companions having $R \sim 50 \km$ ({\it Weaver et
al.}, 2006). The three satellites' orbits are nearly co-planar; their
semi-major axes are about 19600, 48700, and 64800 km; and their eccentricities
are less than 1\% ({\it Buie et al.}, 2006).

\bigskip
\centerline{\textbf{3. THEORETICAL TIMELINE}}
\bigskip

We now recount a possible history of trans-Neptunian space.
Throughout our narration, it is helpful to remember that the timescale
for an object of size $R$ and mass $M$ to collide into its own mass in smaller
objects is

\begin{equation}
t_{\rm col} \sim \frac{M}{\dot{M}} \sim \frac{\rho R^3}{(\sigma / h) R^2 v_{\rm
rel}} \sim \frac{\rho R}{\sigma \Omega} \,,
\label{tcol}
\end{equation}

\ni where $\dot{M}$ is the rate at which mass from the surrounding disk
impacts the object, $\sigma$ is the disk's surface mass density
(mass per unit face-on area) in smaller objects,
$v_{\rm rel}$ is the relative collision velocity,
$h \sim v_{\rm rel}/\Omega$ is the effective vertical scale height
occupied by colliders, and $\Omega$ is the orbital angular frequency.
Relative velocities $v_{\rm rel}$ depend on how $e$ and $i$ are distributed.
Equation (\ref{tcol})
requires that $e$'s and $i$'s be comparably distributed and large enough that
gravitational
focussing is ignorable.
While these conditions are largely met by currently observed KBOs, they
were not during the primordial era. Our expressions below
represent appropriate modifications of (\ref{tcol}).

\bigskip
\centerline{\textbf{3.1. Coagulation}}
\bigskip

The mass inferred for the present-day Kuiper belt, $\sim$0.05--0.3 $M_{\oplus}$
(\S 2.2),
is well below that thought to have been present while KBOs coagulated.
{\it Kenyon and Luu} (1998, 1999) and {\it Kenyon} (2002), in a
series of particle-in-a-box accretion simulations, find that
$\sim$3--30 $M_{\oplus}$ of primordial solids, spread over
an annulus extending from 32 to 38 AU, are required
to (a) coagulate at least 1 object as large as Pluto and (b) coagulate
$\sim$$10^5$ objects having $R > 50\km$.
The required initial surface density,
$\sigma \sim 0.06$--$0.6 \gm \cm^{-2}$, is of order that
of the condensible portion of the minimum-mass solar nebula (MMSN)
at 35 AU: $\sigma_{\rm MMSN} \sim 0.2 \gm \cm^{-2}$.

\bigskip
\centerline{{3.1.1. The Missing-Mass Problem}}
\bigskip

That primordial and present-day masses differ by 2 orders of magnitude
is referred to as the ``missing-mass'' problem.
The same accretion simulations point to a possible resolution:
Only $\sim$1--2\% of the primordial mass accretes to sizes
exceeding $\sim$100 km. The remainder stalls at comet-like sizes of
$\sim$0.1--10 km. Stunting of accretion is attributed
to the formation of several Pluto-sized objects whose
gravity amplifies velocity dispersions so much
that collisions between planetesimals are erosive rather than
accretionary.
Thus, accretion in the Kuiper belt may be
self-limiting ({\it Kenyon and Luu}, 1999).
The bulk of the primordial mass, stalled at cometary sizes,
is assumed by these authors to erode away by destructive collisions over Gyr
timescales.

We can verify analytically some of {\it Kenyon and Luu}'s results
by exercising the ``two-groups method'' ({\it G}04), whereby the
spectrum of planetesimal masses is approximated as bimodal.
``Big'' bodies each of size $R$, mass $M$, 
Hill radius $R_{\rm H}$, and surface escape velocity $v_{\rm esc}$
comprise a disk of surface density $\Sigma$.
They are held primarily responsible for stirring and accreting ``small''
bodies of size $s$, surface density $\sigma$, and random velocity dispersion
$u$.
By random velocity we mean the non-circular or non-planar component of the
orbital velocity.
As such, $u$ is proportional to the root-mean-squared dispersion in $e$ and
$i$.

To grow a big body takes time

\begin{equation}
t_{\rm acc} \equiv \frac{R}{\dot R} \sim \frac{\rho R}{\sigma \Omega} \left(
\frac{u}{v_{\rm esc}} \right)^2 \,,
\label{rrdot}
\end{equation}

\ni where the term in parentheses is the usual gravitational focussing factor
(assumed $< 1$). Gravitational
stirring of small bodies by big ones
balances damping of relative velocities by inelastic collisions amongst small
bodies. This balance sets the equilibrium velocity dispersion $u$:

\begin{equation}
\frac{\rho R}{\Sigma \Omega} \left( \frac{u}{v_{\rm esc}} \right)^4 \sim
\frac{\rho s}{\sigma \Omega} \,.
\label{vsie}
\end{equation}

\ni 
Combining (\ref{rrdot}) and (\ref{vsie}) implies

\begin{eqnarray}
t_{\rm acc} \sim 10 \left( \frac{R}{100 \km} \, \frac{s}{100 \m} \,
\frac{\Sigma/\sigma}{0.01} \right)^{1/2} \left( \frac{\sigma_{\rm
MMSN}}{\sigma} \right) \Myr \\
u \sim 6 \left( \frac{s}{100 \m} \, \frac{\Sigma/\sigma}{0.01} \right)^{1/4}
\left( \frac{R}{100 \km} \right)^{3/4} \m \s^{-1} 
\end{eqnarray}

\ni at a distance of 35 AU. All bodies reside in a remarkably thin, dynamically
cold disk:
Eccentricities and inclinations are at most of order
$u/(\Omega a) \sim 0.001$. Our nominal choices for $\sigma$, $\Sigma$, and
$s$ are informed by {\it Kenyon and Luu}'s proposed solution to the
missing-mass
problem. Had we chosen values
resembling those of the Kuiper belt today---$\Sigma \sim \sigma \sim 0.01
\sigma_{\rm MMSN}$---coagulation times would exceed the age of the solar
system.

The above framework for understanding the missing-mass problem,
while promising, requires development.
First, account needs to be made for how the formation of
Neptune---and possibly other planet-sized bodies---influences
the coagulation of KBOs. None of the simulations cited above
succeeds in producing Neptune-mass objects.
Yet minimum-mass disks may be capable,
in theory if not yet in simulation, of producing several planets
having masses approaching that of Neptune at distances of 15--25 AU on
timescales
much shorter than the age of the solar system ({\it G}04; see
\S\S3.4--3.5).
The inability of simulations to produce
ice giants may arise from their neglect of small-$s$,
low-$u$ particles that can be efficiently accreted ({\it G}04).
Sizes $s$ as small as centimeters seem possible.
How would their inclusion, and the consequent formation
of Neptune-mass planets near the Kuiper belt, change our understanding
of the missing-mass problem?
Second, how does the outer solar system shed $\sim$99\% of its
primordial solids?
The missing-mass problem translates to a ``clean-up'' problem, the solution
to which will involve some as yet unknown combination of collisional
comminution,
diffusive transport by interparticle collisions,
gravitational ejection by planets, and removal by gas and/or radiation drag.

\bigskip
\centerline{{3.1.2. The Outer Edge of the Primordial Planetesimal Disk}}
\bigskip

How far from the Sun did planetesimals coagulate? The outer edge
of the Classical disk at 47 AU (\S 2) suggests that planetesimals failed
to form outside this distance. Extra-solar disks are also observed to have
well-defined boundaries.
The debris disks encircling $\beta$ Pictoris
and AU Microscopii exhibit distinct changes (``breaks'') in the slopes of their
surface
brightness profiles at stellocentric distances of $100 \AU$ and
$43 \AU$, respectively (e.g., {\it Kalas and Jewitt}, 1995; {\it Krist et al.},
2005). This behavior can be explained
by having dust-producing parent bodies reside only at distances interior to
break radii ({\it Strubbe and Chiang}, 2006).

We cannot predict with confidence how planetesimal disks truncate.
Our understanding of how micron-sized dust assembles into the ``small,''
super-meter-sized bodies that coagulation calculations presume as input is too
poor.
Recent work discusses how solid particles might drain 
towards their central stars by gas drag, and how the accumulation of such
solids at small stellocentric
distances triggers self-gravitational collapse and
the formation of larger bodies ({\it Youdin and Shu}, 2002; {\it Youdin and
Chiang}, 2004).
These ideas promise to explain why planetesimal disks have sharp
outer edges, but are subject to uncertainties regarding the viability
of gravitational instability in a turbulent gas. To sample progress
on planetesimal formation,
see {\it Garaud and Lin} (2004), {\it Youdin and Goodman} (2005),
and {\it G\'omez and Ostriker} (2005). In what follows, we assume
that objects having $R \sim 100 \km$ coagulated only
inside 47 AU.

\bigskip
\centerline{\textbf{3.2. Formation of Binaries}}
\bigskip

To have formed from a fragmentary collision, binary components
observed today cannot have too much angular momentum.
Consider two big bodies undergoing a gravitationally focussed
collision. Each body has radius $R$, mass $M$, and surface escape
velocity $v_{\rm esc}$. Prior to the collision, their angular momentum is at
most
$L_{\rm max} \sim Mv_{\rm esc}R$.
After the collision, the resultant binary must have angular momentum
$L < L_{\rm max}$. Unless significant mass is lost from the collision,
components can be comparably sized
only if their separation is comparable to their
radii. Pluto and Charon meet this constraint.
Their mass ratio is $\sim$$1/10$, their separation is $\sim$$20 R_{\rm
Pluto}$,
and hence their angular momentum obeys $L/L_{\rm max} \sim \sqrt{20}/10
\lesssim 1$.
{\it Canup} (2005) explains how Charon might have formed by
a collision. The remaining satellites of Pluto ({\it Stern et al.}, 2006), the
satellites of Pluto-sized KBOs 2003 EL$_{61}$
and 2003 UB$_{313}$, and the candidate near-contact binaries
discovered by {\it Sheppard and Jewitt} (2004) might also have formed by
collisions.

By contrast, binary components having wide separations and comparable masses
have too much angular momentum to have formed by gravitationally focussed
collisions.
And if collisions were unfocussed, collision times
would exceed the age of the solar system---assuming, as we do
throughout this review, that the surface density of big ($R \sim 100\km$)
bodies
was the same then as now (\S1; \S3.9).

Big bodies can instead become bound (``fuse'') by purely gravitational means
while they are still dynamically cold.
Indeed, such binaries testify powerfully to the cold state
of the primordial disk. To derive our expressions below, recall that
binary components separated by the Hill radius $\sim$$R_{\rm H}$ orbit each
other
with the same period that the binary's center of mass orbits the Sun,
$\sim$$\Omega^{-1}$. Furthermore, we assume that the velocity dispersion $v$
of big bodies is less than their Hill velocity $v_{\rm H} \equiv \Omega R_{\rm
H}$.
Then big bodies undergo runaway cooling by dynamical friction with small bodies
and
settle into an effectively two-dimensional disk ({\it G}04). Reaction rates
between big bodies
must be calculated in a 2-D geometry. Because
$u > v_{\rm H}$, reaction rates involving small bodies take
their usual forms appropriate for three dimensions.

{\it Goldreich et al.}~(2002, hereafter {\it G}02) describe two collisionless
formation scenarios, dubbed $L^3$ and $L^2s$.
Both begin when one big body ($L$) enters a second big body's ($L$) Hill
sphere.
Per big body, the entry rate is

\begin{equation}
\dot{N}_{\rm H} \sim \frac{\Sigma \Omega}{\rho R} \left( \frac{R_{\rm H}}{R}
\right)^2 \sim \frac{\Sigma \Omega}{\rho R} \alpha^{-2} \,,
\label{eq:enter}
\end{equation}

\ni where $\alpha \equiv R/R_{\rm H} \approx 1.5 \times 10^{-4} (35 \AU / a)$.
If no other body participates in the interaction,
the two big bodies would pass through their Hill spheres in a time
$\sim$$\Omega^{-1}$ (assuming they do not collide).
The two bodies fuse if they transfer enough energy to other participants
during the encounter.
In $L^3$, transfer is to a third big body: $L+L+L\rightarrow L^2+L$.
To just bind the original pair,
the third body must come within $R_{\rm H}$ of the pair.
The probability for this to happen in time $\Omega^{-1}$ is
$P_{L^3} \sim \dot{N}_{\rm H} \Omega^{-1}$. If the third
body succeeds in approaching this close, the probability
that two bodies fuse is order unity. Therefore the timescale
for a given big body to fuse to another by $L^3$ is

\begin{equation}
t_{{\rm fuse}, L^3} \sim \frac{1}{\dot{N}_{\rm H} P_{L^3}} \sim \left(
\frac{\rho R}{\Sigma} \right)^2 \frac{\alpha^4}{\Omega} \sim 2 \Myr \,,
\end{equation}

\ni where the numerical estimate assumes $R=100$ km, $\Sigma=0.01\sigma_{\rm
MMSN}$, and $a = 35 \AU$.

In $L^2s$, energy transfer is to small bodies by dynamical friction:
$L+L+s^\infty\rightarrow L^2+s^\infty$.
In time $\Omega^{-1}$, the pair of big bodies undergoing an encounter
lose a fraction $(\sigma \Omega/ \rho R) (v_{\rm esc}/u)^4 \Omega^{-1}$
of their energy, under the
assumption $v_{\rm esc} > u > v_{\rm H}$ ({\it G}04).
This fraction is of order the probability
$P_{L^2s}$ that they fuse, whence

\begin{equation}
t_{{\rm fuse},L^2s} \sim \frac{1}{\dot{N}_{\rm H} P_{L^2s}} \sim \left({\rho R
\over\sigma}\right)^{2} {s\over R} \frac{\alpha^{2}}{\Omega} \sim 7 \Myr \,,
\end{equation}

\ni where we have used (\ref{vsie}) and set $s = 100$ m.

Having formed with semi-major axis $x \sim R_H \sim 7000 R$,
a binary's orbit shrinks by further energy transfer.
If $L^3$ is the more efficient formation
process, passing big bodies predominantly harden the binary;
if $L^2s$ is more efficient, dynamical friction dominates hardening.
The probability $P$ per orbit that $x$
shrinks from $\sim$$R_{\rm H}$ to $\sim$$R_{\rm H}/2$
is of order either $P_{L^3}$ or $P_{L^2s}$.
We equate the formation rate of binaries, $N_{\rm all}/t_{\rm fuse}$,
with the shrinkage rate, $\Omega P N_{{\rm bin}}|_{x \sim R_{\rm H}}$,
to conclude that the steady-state fraction of KBOs that are
binaries with separation $R_{\rm H}$ is

\begin{equation}
f_{\rm bin} (x \sim R_{\rm H}) \equiv \left. \frac{N_{\rm bin}}{N_{\rm all}}
\right|_{x \sim R_{\rm H}} \sim \frac{\Sigma}{\rho R} \alpha^{-2} \sim 0.4\%
\,.
\end{equation}

\ni As $x$ decreases, shrinkage slows. Therefore $f_{\rm bin}$
increases with decreasing $x$. Scaling relations can be derived
by arguments similar to those above. If $L^2s$ dominates,
$f_{\rm bin} \propto x^0$ for $x > R_{\rm H} (v_{\rm H}/u)^2$
and $f_{\rm bin} \propto x^{-1}$ for $x < R_{\rm H} (v_{\rm H}/u)^2$ ({\it
G}02).
If $L^3$ dominates,  $f_{\rm bin} \propto x^{-1/2}$.

Alternative formation scenarios require, in addition to gravitational
scatterings, physical collisions. {\it Weidenschilling} (2002) suggests
a variant of $L^3$ in which the third big body collides with one member
of the scattering pair. Since physical collisions
have smaller cross-sections than gravitational interactions,
this mechanism requires $\sim$$10^2$ more big ($R \sim 100\km$) bodies
than are currently observed
to produce the same rate that is cited above for $L^3$
({\it Weidenschilling}, 2002).
{\it Funato et al.}~(2004) propose that observed binaries form by the
exchange reaction $Ls+L\rightarrow L^2 +s$: A small body of mass $m$,
originally orbiting a big body of mass $M$, is ejected by a
second big body. In the majority of ejections, the small
body's energy increases by its orbital binding energy
$\sim$$mv_{\rm esc}^2/2$, leaving the big bodies bound
to each other with separation $x \sim (M/m)R$.
The rate-limiting step is the formation
of the pre-existing ($Ls$) binary, which requires (as in the asteroid belt)
two big bodies to collide and fragment. Hence

\begin{equation}
t_{{\rm fuse,exchange}} \sim \frac{\rho R}{\Sigma \Omega} \alpha^{3/2} \sim 0.6
\Myr \,.
\end{equation}

\ni 
Estimating $f_{\rm bin}$ as a function of $x$ under the exchange
hypothesis requires knowing the distribution of fragment masses $m$.
Whether $L^2s$, $L^3$, or exchange reactions dominate depends on
the uncertain parameters
$\Sigma$, $\sigma$, and $s$.

As depicted above, newly formed binaries
should be nearly co-planar with the 2-D disk of big bodies,
i.e., binary orbit normals should be nearly parallel.
Observations contradict this picture (\S 2.3).
How dynamical stirring of the Kuiper belt subsequent to binary formation
affects binary inclinations and eccentricities has not been investigated.

\bigskip
\centerline{\textbf{3.3. Early Stirring by Growing Planetary Oligarchs}}
\bigskip

Coagulation of KBOs and fusing of binaries cannot proceed today,
in part because velocity dispersions are now so large
that gravitational focussing is defeated on a wide range of length scales.
What stirred the Kuiper belt?
There is no shortage of proposed answers.
Much of the remaining review (\S\S 3.3--3.7)
explores the multitude of non-exclusive
possibilities. We focus on stirring ``large'' KBOs like those
currently observed, having $R \sim 100 \km$. 
Our setting remains the primordial
disk, of whose mass large KBOs constitute only a small fraction (1--2\%;
\S3.1.1).

 
Neptune and Uranus are thought to accrete as oligarchs,
each dominating their own annulus of full width $\sim$5 Hill radii ({\it G}04;
{\it Ida and Makino}, 1993; {\it Greenberg et al.}, 1991;
the coefficient of 5 presumes that oligarchs
feed in a shear-dominated disk in which
planetesimals
have random velocities
$u$ that are less than the oligarch's Hill velocity $v_{\rm H} = \Omega R_{\rm
H}$.
If $u > v_{\rm H}$, oligarchs' feeding
annuli are wider by $\sim$$u/v_{\rm H}$. In practice, $u/v_{\rm H}$ does not
greatly exceed unity
since it scales weakly with input parameters.)
Each oligarch grows until its mass equals the isolation mass,

\begin{equation}
M_{\rm p} \sim 2 \pi a \times 5 R_{\rm H,p} \times \sigma \,,
\label{isolation}
\end{equation}

\ni where $R_{\rm H,p}$ is the oligarch's Hill radius.
For $a = 25 \AU$ and $M_{\rm p}$ equal to Neptune's mass $M_{\rm N} = 17
M_{\oplus}$,
equation (\ref{isolation}) implies
$\sigma \sim 0.9 \gm \cm^{-2} \sim 3 \sigma_{\rm MMSN}$.
About 5 Neptune-mass oligarchs can form in nested annuli
between 15 and 25 AU. Inspired by {\it G}04 who point out
the ease with which ice giants coagulate when the bulk of the disk
mass comprises very small particles (\S 3.1.1),
we assume that all 5 do form in a disk that is a few times more massive
than the MMSN and explore the consequences of such an initially packed system.

While oligarchs grow, they stir large KBOs in their immediate vicinity. 
A KBO that comes within distance $b$
of mass $M_{\rm p}$ has its random velocity
excited to $v_{\rm K} \sim (G M_{\rm p}/b)^{1/2}$.
Take the surface density of perturbers to be $\Sigma_{\rm p}$. Over time $t$, a
KBO
comes within distance $b \sim [M_{\rm p} / (\Sigma_{\rm p} \Omega t)]^{1/2}$
of a perturber. Therefore

\begin{equation}
v_{\rm K} \sim G^{1/2} (M_{\rm p} \Sigma_{\rm p} \Omega t)^{1/4} \,.
\label{vkt}
\end{equation}

\noindent Since Neptune and Uranus contain more hydrogen than can
be explained by accretion of icy solids alone, they must complete
their growth within $t_{\rm acc,p} \sim 1$--10 Myr, before all hydrogen gas 
in the MMSN photo-evaporates (e.g., {\it Matsuyama et al.}, 2003, and
references therein).
For $t = t_{\rm acc,p} = 10 \Myr$, $M_{\rm p} = M_{\rm N}$,
$\Sigma_{\rm p} = 0.9 \gm \cm^{-2}$, and $\Omega = 2\pi/(100 \yr)$,
equation (\ref{vkt}) implies $v_{\rm K} \sim 1 \km \s^{-1}$
or $e_{\rm K} \sim 0.2$.
It is safe to neglect damping of $v_{\rm K}$ for large KBOs, which occurs
by inelastic collisions over a timescale
$t_{\rm col} \sim 400 \,(0.9 \gm \cm^{-2} / \sigma)\,\Myr \gg t_{\rm acc,p}$.

\bigskip
\centerline{\textbf{3.4. Velocity Instability and Ejection of Planets}}
\bigskip

Once the cohort of Neptune-mass oligarchs consumes $\sim$1/2 the mass of the
parent disk, they
scatter one another onto highly elliptical and inclined orbits ({\it G}04, see
their equation [111]; {\it Kenyon and Bromley}, 2006).
This velocity instability occurs because damping of planetary random velocities
by dynamical friction with the disk can no longer compete
with excitation by neighboring, crowded oligarchs.

The epoch of large planetary eccentricities lasts until enough oligarchs
are ejected from the system. We can estimate the ejection time
by following the same reasoning that led to (\ref{vkt}).
Replace $v_{\rm K}$ with the system escape velocity
$v_{\rm esc,sys} \sim \Omega a$, and replace $\Sigma_{\rm p}$ with the surface
density of oligarchs $\sim$$M_{\rm p} / a^2$ (see equation [\ref{isolation}]).
Then solve for

\begin{equation}
t = t_{\rm eject} \sim \left( \frac{M_{\odot}}{M_{\rm p}} \right)^2
\frac{0.1}{\Omega} \,.
\label{teject}
\end{equation}


\noindent The coefficient of 0.1 is attributed to more careful accounting
of encounter geometries; equation (\ref{teject}) gives ejection times
similar to those found in numerical simulations ({\it G}04). 
Neptune-mass oligarchs at $a \approx 20 \AU$ 
kick their excess brethren out
over $t_{\rm eject} \sim 600 \Myr$.
Removal is faster if
excess oligarchs are passed inward to Jupiter and Saturn.

Oligarchs moving on eccentric orbits likely traverse distances
beyond 30 AU and stir KBOs.
We expect more members are added to the Scattered KBO disk during this stage.

We have painted a picture of dynamically hot oligarchs
similar to that drawn by {\it Thommes et al.}~(1999; see also {\it Tsiganis et
al.}, 2005),
who hypothesize that Neptune and Uranus form as oligarchs
situated between the cores of Jupiter and Saturn at 5--10 AU.
The nascent ice giants are scattered outward onto eccentric
orbits once the gas giant cores amass their envelopes.
While Neptune and Uranus reside on eccentric orbits, they can stir
KBOs in much the same way as we have described above ({\it Thommes et al.},
2002).
Despite the similarity of implications for the stirring of KBOs,
the underlying motivation of the cosmogony proposed by {\it Thommes et
al.}~(1999)
is the belief that Neptune-mass bodies
do not form readily at distances of $\sim$30 AU.
Recent work highlighting the importance of inelastic collisions amongst very
small
bodies challenges this belief ({\it G}04; see \S3.1).

\bigskip
\centerline{\textbf{3.5. Dynamical Friction Cooling of Surviving Planets}}
\bigskip

Planetary oligarchs that survive ejection---i.e., Uranus and Neptune---have
their $e$'s and $i$'s restored
to small values by dynamical friction with the remnant disk (comprising
predominantly small KBOs of surface density $\sigma$ and velocity dispersion
$u$)
over time

\begin{equation}
t_{\rm df,cool} = \frac{v_{\rm p}}{\dot{v}_{\rm p}} \sim \frac{\rho R_{\rm
p}}{\sigma \Omega} \left( \frac{v_{\rm p}}{v_{\rm esc,p}} \right)^4 \,,
\label{dfcool}
\end{equation}

\ni where $R_{\rm p}$, $v_{\rm esc,p}$, and
$v_{\rm p} \gg u$ are the planet's
radius, surface escape velocity, and random velocity, respectively.
For $v_{\rm p} = \Omega a/2$ (planetary eccentricity
$e_{\rm p} \sim 0.5$),
$a = 25 \AU$,
$R_{\rm p} = 25000 \km$,
$v_{\rm esc,p} = 24 \km \s^{-1}$,
and $\sigma = \Sigma_{\rm p} = 0.9 \gm \cm^{-2}$
(since the velocity instability occurs when the surface density
of oligarchs equals that of the parent disk; \S3.4),
we find $t_{\rm df,cool} \sim 20\Myr$.

While Neptune's orbit is eccentric, the planet might repeatedly invade
the Kuiper belt at $a \approx 40$--45 AU and stir KBOs.
Neptune would have its orbit circularized by transferring energy to both small
and large KBOs.
Unlike small KBOs, large ones cannot shed
this energy because they cool too inefficiently by inelastic collisions
(see the end of \S3.3).
Insert (\ref{dfcool}) into (\ref{vkt}) and set $\Sigma_{\rm p} = \sigma$ to
estimate the random velocity to which large KBOs are excited by a cooling
Neptune:

\begin{equation}
v_{\rm K} \sim v_{\rm p} \,.
\label{vkvp}
\end{equation}

\ni Thus large KBOs are stirred to the same random velocity that Neptune had
when the
latter began to cool, regardless of the numerical value of $t_{\rm df,cool}$.
Large KBOs effectively record the eccentricity of Neptune just prior to its
cooling phase.
Final eccentricities $e_{\rm K}$ might range
from $\sim$0.1 to nearly 1. During this phase, the population
of the Scattered KBO disk would increase, perhaps dramatically so.
If all large KBOs are stirred to $e_{\rm K} \gg 0.1$, new large
KBOs must coagulate afterwards from the
remnant disk of small, dynamically cold bodies
to re-constitute the cold Classical disk. Cold Classicals
might therefore post-date hot KBOs.

\bigskip
\centerline{\textbf{3.6. Planetary Migration}}
\bigskip

Having seen a few of its siblings evicted, and having settled
onto a near-circular, flattened orbit, Neptune remains
immersed in a disk of small bodies. 
The total mass of the disk
is still a few times that of the planet because the prior velocity instability
occurred when
the surface density of oligarchs was comparable to that of the disk.
By continuing to scatter small bodies,
Neptune migrates: Its semi-major axis changes while
its eccentricity is kept small by dynamical friction. Absent other planets,
migration would be Sunward on average. 
Planetesimals repeatedly scattered by Neptune would exchange angular momentum
with
the planet in a random-walk fashion. Upon gaining specific angular momentum
$\sim$$(\sqrt{2}-1)$$\Omega a^2$, where $\Omega$ and $a$ are appropriate
to Neptune's orbit, a planetesimal initially near Neptune
would finally escape. Having lost angular momentum to the ejected
planetesimal,
Neptune would migrate inward. (A single planet can still migrate outward if it
scatters material having predominantly higher specific angular momentum.
{\it Gomes et al.}~(2004) achieve this situation by embedding Neptune
in a disk whose mass is at least $100 M_{\oplus}$
and is weighted towards large distances [$\sigma a^2 \propto a$]; see also the
chapter
by {\it Morbidelli et al.}).

Other planets complicate this process.
Numerical simulations by {\it Fern\'andez and Ip}
(1984) and {\it Hahn and Malhotra} (1999) incorporating all 4 giant planets
reveal that planetesimals
that originate near Neptune are more likely ejected by Jupiter.
Over the course of their random walks, planetesimals lose angular momentum to
Neptune
and thereby cross Jupiter's orbit. Jupiter summarily ejects
them (see equation [\ref{teject}] and related discussion).
Thus, on average, Neptune gains angular momentum and migrates outward,
as do Saturn and Uranus, while Jupiter's orbit shrinks.

An outward bound Neptune passes objects to the interior planets
for eventual ejection and seeding of the Oort Cloud.
We refer to this process as ``scouring'' the trans-Neptunian disk.
Scouring and migration go hand in hand; the fraction by which Neptune's
semi-major axis increases is of order the fraction that the disk
mass is scoured.
Scouring is likely a key part of the solution to the clean-up
(a.k.a.~missing-mass) problem.
If clean-up is not achieved by the end of Neptune's migration, one must
explain
how to transport the bulk of the trans-Neptunian disk to other locales
while keeping Neptune in place ({\it Gomes et al.}, 2004).
Scouring has only been treated in collisionless N-body simulations.
How scouring and migration proceed in a highly collisional disk
of small bodies is unknown (\S 3.6.4).

In addition to scouring the disk, Neptune's migration
has been proposed to sculpt the disk in other ways---by capturing
bodies into mean-motion resonances (\S 3.6.1), re-distributing
the Classical disk by resonance capture and release (\S 3.6.2),
and deflecting objects onto Scattered orbits (\S 3.6.3).
We critically examine these proposals below.

\bigskip
\centerline{{3.6.1. Capture and Excitation of Resonant KBOs}}
\bigskip

As Neptune migrates outward, its exterior mean-motion resonances (MMRs)
sweep across trans-Neptunian space. Provided the migration
is sufficiently slow and smooth,
MMRs may trap KBOs and amplify their orbital eccentricities
and, to a lesser extent, their inclinations. The eccentric orbits of Pluto and
the Plutinos---objects
which all inhabit Neptune's 3:2 resonance---may have resulted
from resonance capture and excitation by a migrating Neptune ({\it Malhotra},
1993, 1995;
{\it Jewitt and Luu}, 2000). The observed occupation of other low-order
resonances---e.g., the 4:3, 5:3, and 2:1 MMRs---by
KBOs on eccentric orbits (see Fig.~2 and Table 1) further support the migration
hypothesis
({\it C}03). In this section,
we review the basic mechanism of resonant
excitation of eccentricity, examine how the migration hypothesis must
change in light of the unexpected occupation of high-order (e.g., the 7:4, 5:2,
and
3:1) MMRs, and discuss how $m$:1 resonances serve as speedometers
for Neptune's migration.

Consider the interaction between a test particle (KBO)
and a planet on an expanding circular orbit.
In a frame of reference centered on the Sun and rotating with the planet's
angular velocity $\Omega_{\rm p}(t)$, the particle's Hamiltonian is

\begin{equation}\label{ham}
\mathcal{H} = \mathcal{E}-\Omega_{\rm p}(t)\mathcal{L} - \mathcal{R}(t) \,,
\end{equation}

\ni where 
$\mathcal{E} = -GM_{\odot}/2a$,
$\mathcal{L} = [GM_{\odot}a(1-e^2)]^{1/2}$, and
$\mathcal{R}$ is the disturbing potential due to the planet
(these quantities
should be expressed in
canonical coordinates).
{}From Hamiltonian mechanics,
$d\mathcal{H}/dt = \partial \mathcal{H}/\partial t = - \dot{\Omega}_{\rm p}
\mathcal{L} - \partial{\mathcal{R}}/\partial{t}$.
Therefore 

\begin{equation}\label{fromhamdot}
\frac{d\mathcal{E}}{dt}(1-\epsilon) - \Omega_{\rm p} \frac{d\mathcal{L}}{dt} =
0 \,, 
\end{equation}

\ni where $\epsilon\equiv (d\mathcal{R}/dt-\partial \mathcal{R}/\partial
t)/(d\mathcal{E}/dt)$.  We re-write (\ref{fromhamdot}):

\begin{equation}
\frac{de^2}{dt} = \frac{ (1-e^2)^{1/2}}{a} \left[ (1-e^2)^{1/2} -
\Omega/\Omega_{\rm p}(1-\epsilon) \right] \frac{da}{dt}\,,
\label{jdot}
\end{equation}

\ni where $\Omega$ is the particle's angular frequency.

For a particle trapped in $m$:$n$ resonance
(where $m$ and $n$ are positive, relatively prime integers),
$a$, $e$, and the resonance angle change little over the particle's
orbital period.  If the synodic period is not much longer than the orbital
period, we may average the Hamiltonian over the former (we may do
this by choosing appropriate terms in the expansion of $\mathcal{R}$).
This yields $\Omega/\Omega_{\rm p}(1-\epsilon) = n/m$.
For a particle in resonance, $|\epsilon| \ll 1$.
By change of variable to $x \equiv (1-e^2)^{1/2}$,
equation (\ref{jdot}) integrates to

\begin{equation}
\left[ (1-e^2)^{1/2} - n/m \right]^2 a = {\rm constant} \,,
\label{brouwer}
\end{equation}

\ni which relates changes in $a$ to changes in $e$
for any resonance---exterior $m > n$, interior $m < n$, or Trojan $m = n$.
In the case of a planet that migrates towards a particle in exterior
resonance,
$a$ increases to maintain resonant lock ({\it Goldreich}, 1965; {\it Peale},
1986).
Then by (\ref{brouwer}), $e$ also tends to increase, towards a maximum value
$[1-(n/m)^2]^{1/2}$.
Particles inhabiting either an exterior or interior resonance have their
eccentricities
amplified from 0 because they are perturbed by a force pattern whose angular
speed $\Omega_{\rm p}$
does not equal their orbital angular speed $\Omega$. Particles
receive energy and angular momentum from the planet in a ratio
that cannot maintain circularity of orbits.

Among observed 2:1 Resonant KBOs, $\max (e) \approx 0.38$ (Fig.~2).
If 2:1 Resonant KBOs had their eccentricities amplified purely by
migration, they must have migrated by $\Delta a \approx 13 \AU$ (equation
[\ref{brouwer}]).
Neptune must have migrated correspondingly by
$\Delta a_{\rm p} \approx 8 \AU$. This is an upper bound
on $\Delta a_{\rm p}$ because it does not account for non-zero initial
eccentricities prior to capture.

In early simulations ({\it Malhotra}, 1993, 1995) of resonance capture by a
migrating Neptune,
resonances swept across KBOs having initially small $e$'s
and $i$'s. These models predicted that if Neptune's orbit expanded
by $\Delta a_{\rm p} \approx 8 \AU$, low-order resonances
such as the 4:3, 3:2, 5:3, and 2:1 MMRs would be occupied by
objects having $0.1 \lesssim e \lesssim 0.4$ and $i \lesssim 10^{\circ}$.
Eccentric KBOs indeed inhabit these resonances (Fig.~2).
Two observations were not anticipated:
(1) Resonant KBOs are inclined by up to $i \approx 30^{\circ}$, and (2)
high-order
resonances---e.g., the 5:2, 7:4, and 3:1---enjoy occupation.
These observations suggest that Neptune's MMRs swept
across not only initially dynamically cold objects, but also initially hot
ones:
The belt was pre-heated.
For example, to capture KBOs into the 5:2 MMR, pre-heated eccentricities
must be $\gtrsim$ 0.1 ({\it C}03). Neptune-sized perturbers (\S 3.3--3.5)
might have provided the requisite pre-heating in $e$ and $i$.

To understand why capture into high-order resonances favors particles having
larger initial $e$,
recognize that capture is only possible if, over the time
the planet takes to migrate across the maximum possible libration width $\max
(\delta a_{\rm lib})$, 
the particle completes at least 1 libration:

\begin{equation}
\frac{\max (\delta a_{\rm lib})}{ |\dot{a}_{\rm p}| } > T_{\rm lib} \,,
\label{condition}
\end{equation}

\ni where $T_{\rm lib}$ is the libration period
({\it Dermott et al.}, 1988). Otherwise, the particle would hardly feel
the resonant perturbation
as the planet races towards it. Since $\max (\delta a_{\rm lib}) \sim (T_{\rm
orb}/T_{\rm lib}) a_{\rm p}$
and $T_{\rm lib} \sim T_{\rm orb} (M_{\odot} e^{-|m-n|}/M_{\rm p})^{1/2}$ ({\it
Murray and Dermott}, 1999),
where $T_{\rm orb}$ is the orbital period of the particle and $M_{\rm p}$ is
the
mass of the planet, we re-write equation (\ref{condition}) as

\begin{equation}
\frac{T_{\rm lib}^2}{T_{\rm orb} T_{\rm mig}} \sim \frac{T_{\rm orb}}{T_{\rm
mig}} \frac{M_{\odot}}{M_{\rm p}} \frac{1}{e^{|m-n|}} < 1 \,,
\label{condition2}
\end{equation}

\ni where the migration timescale $T_{\rm mig} \equiv a_{\rm p}/|\dot{a}_{\rm
p}|$.
The higher the order $|m-n|$ of the resonance,
the greater $e$ must be to satisfy (\ref{condition2}) ({\it C}03; {\it Hahn and
Malhotra}, 2005).

Asymmetric ($m$:1) resonances afford a way to estimate the migration timescale
observationally.
An asymmetric MMR furnishes multiple islands of libration.
At the fixed point of each island, a particle's direct acceleration by Neptune
balances its indirect acceleration by the Sun due to the Sun's
reflex motion ({\it Pan and Sari}, 2004; {\it Murray-Clay and Chiang}, 2005,
hereafter {\it MC}05).
The multiplicity of islands translates into a multiplicity of orbital
longitudes, measured relative
to Neptune's, where resonant KBOs cluster on the sky. The pattern of
clustering
varies systematically with migration speed at the time of capture ({\it Chiang
and Jordan}, 2002).
For example, when migration is fast---occurring on timescales
$T_{\rm mig} \lesssim 20 \Myr$---objects are caught into 2:1 resonance
such that more appear at longitudes trailing, rather than leading, Neptune's.
The degree of asymmetry can be as large as 300\%.  When migration
is slow, the distribution of captured 2:1 objects is symmetric about the
Sun-Neptune line. The preference for trailing versus
leading longitudes arises from migration-induced shifts in the stable and
unstable equilibria
of the resonant potential. Shifts in the equilibrium values
of the resonance angle are given in radians by
equation (\ref{condition2}) and are analogous to the shift in the equilibrium
position of a spring in a gravitational field ({\it MC}05).
The observation that trailing 2:1 KBOs do not outnumber leading ones
constrains $T_{\rm mig} > 20 \Myr$ with nearly $3\sigma$ confidence ({\it
MC}05).
This measurement accords with numerical simulations of the migration
process itself by {\it Hahn and Malhotra} (1999) and by {\it Gomes et
al.}~(2004, see their fig.~10);
in these simulations, $T_{\rm mig} \gtrsim 40 \Myr$.

\bigskip
\centerline{3.6.2. Stochastic Migration and Resonance Retainment}
\bigskip

Finite sizes of planetesimals render planetary migration stochastic
(``noisy'').
The numbers of high and low-momentum objects
that Neptune encounters over fixed time intervals fluctuate randomly.
These fluctuations sporadically hasten and slow---and might occasionally
even reverse---the planet's migration.
Apportioning a fixed disk mass to larger (fewer) planetesimals
generates more noise. Extreme noise defeats resonance capture.
Therefore the existence of Resonant KBOs---which we take to imply
capture efficiencies of order unity---sets an upper
limit on the sizes of planetesimals (small bodies) comprising
the bulk of the mass of the disk. {\it Murray-Clay and Chiang} (2006, hereafter
{\it MC}06)
estimate this upper limit to be $s_{\rm max} \sim \mathcal{O}(100) \km$; a
shortened derivation of their result reads as follows.

For a given planetesimal size, most noise is generated per unit mass disk
by planetesimals having sub-Hill ($u < v_{\rm H,p} = \Omega R_{\rm p}/\alpha$)
velocity dispersions
and semi-major axes displaced $\pm R_{\rm H,p}$ from the planet's ({\it
MC}06).
A single such planetesimal of mass $\mu$, after undergoing a close encounter
with the planet, changes the planet's semi-major axis
by $\Delta a_1 \sim \pm (\mu/M_{\rm P}) R_{\rm H,p}$.
The planet encounters such planetesimals at a rate
$\dot{N} \sim \sigma R_{\rm H,p}^2 \Omega_{\rm p} / \mu$.
Over the duration of migration
$\sim$$(\Delta a_{\rm p}/a_{\rm p}) T_{\rm mig}$, 
the planet's semi-major axis random walks away from its nominal (zero-noise)
value by
$\Delta a_{\rm rnd} \sim \pm (\dot{N} \Delta a_{\rm p} T_{\rm mig} / a_{\rm
p})^{1/2} |\Delta a_1|$.
The libration amplitude in $a$ of any resonant KBO
increases by about this same $|\Delta a_{\rm rnd}|$.
Then stochasticity does not defeat
resonance capture if $|\Delta a_{\rm rnd}| < \max (\delta a_{\rm lib})$; that
is, if

\begin{equation}
s \lesssim \left( \frac{M_{\rm p}}{M_{\odot}} \right)^{1/9} \left( \frac{\rho
R_{\rm p} e a_{\rm p}}{\sigma \Omega_{\rm p} T_{\rm mig} \Delta a_{\rm p}}
\right)^{1/3} \alpha^{2/3} R_{\rm p} \,,
\end{equation}

\ni which evaluates to $s \lesssim \mathcal{O}(100) \km$ for $a_{\rm p} =
30\AU$,
$\Delta a_{\rm p} = 8 \AU$, $T_{\rm mig} \approx 40 \Myr$, $\sigma = 0.2 \gm
\cm^{-2}$,
and $e = 0.2$.

The above constraint on size applies to those planetesimals that comprise the
bulk
of the disk mass. Noise is also introduced by especially
large objects that constitute a small fraction of the disk mass.
The latter source of noise has been invoked to explain the
curious near-coincidence
between the edge of the Classical disk ($a$ = 47 AU) and Neptune's 2:1
resonance
($a$ = 47.8 AU). {\it Levison and Morbidelli} (2003) suggest that
the sweeping 2:1 MMR captures KBOs only to release them
en route because of close encounters between Neptune
and objects having $\sim$10 times the mass of Pluto (``super-Plutos'').
Dynamically cold KBOs, assumed to coagulate
wholly inside 35 AU (\S 3.1.2), are thereby combed outwards to fill the space
interior to
the final location of the 2:1 MMR.  Why the super-Plutos that are invoked to
generate
stochasticity have not been detected by wide-field surveys is unclear ({\it
Morbidelli et al.}, 2002).
The scenario further requires
that $\sim$$3 M_{\oplus}$ be trapped within the 2:1 MMR
so that a secular resonance maintains a population of 2:1 resonant KBOs on
low-$e$ orbits
during transport.

\bigskip
\centerline{{3.6.3. Contribution of Migration to Scattered KBOs}}
\bigskip

Neptune migrates by scattering planetesimals.
What fraction of these
still reside today in the Scattered belt?
Do hot Classicals (having $i \gtrsim 5^{\circ}$) owe
their excitation to a migratory Neptune?
Many Scattered and hot Classical
KBOs observed today have $q > 37 \AU$. This fact is difficult
to explain by appealing to perturbers that reside entirely inside 30 AU.
Insofar as a close encounter between a perturber
and a particle can be modelled as a discontinous
change in the particle's velocity at fixed position,
the particle (assuming it remains bound to the Sun)
tends to return to the same location
at which it underwent the encounter.

{\it Gomes} (2003ab) proposes that despite this difficulty,
objects scattered by Neptune during its migration from $\sim$20 to 30 AU
can evolve into today's Scattered and hot Classical KBOs
by having their perihelia raised by a variety of sweeping
secular resonances (SRs; see \S2.1).
As the outer planets migrate, SRs sweep across trans-Neptunian
space. After having its $e$ and $i$ amplified by close
encounters with Neptune, a planetesimal may be swept over by an SR.
Unlike MMRs, SRs cannot alter particle semi-major axes and therefore
do not permanently trap particles. However, a particle that is swept
over by an apsidal-type SR can have its eccentricity
increased or decreased.
A particle swept over by a secular
resonance is analogous to an ideal spring of
natural frequency $\omega_0$, driven
by a force whose time-variable frequency $\omega(t)$ sweeps past
$\omega_0$. Sweeping $\omega$ past $\omega_0$
can increase or decrease the amplitude of the spring's
free oscillation (the component of the spring's displacement
that varies with frequency $\omega_0$), depending
on the relative phasing between driver and spring
near the moment of resonance crossing when $\omega \approx \omega_0$.

Lowering $e$ at fixed $a$ raises $q$.
{\it Gomes} (2003ab) and {\it Gomes et al.}~(2005) find in numerical
simulations of planetary migration
that Neptune-scattered planetesimals originating on orbits inside 28 AU can
have
their perihelia raised up to 69 AU
by a combination of sweeping SRs, MMRs, and Kozai-type resonances (which are a
kind of SR).
In addition to offering an explanation for the origin of high-$q$, high-$i$
KBOs,
this scenario also suggests a framework for understanding differences
in physical properties between dynamical classes.
Compared to Classical KBOs, which are held to coagulate and evolve largely
{\it in situ}, Scattered KBOs originate from smaller heliocentric distances
$d$.
To the (unquantified) extents that coagulation rates and chemical environments
vary
from $d \approx 20$ to 50 AU, we can hope
to understand why a large dispersion in $i$---which in the proposed scenario
reflects a large dispersion in birth distance $d$---implies a large dispersion
in color/size.

The main difficulty with this perihelion-raising mechanism
is its low efficiency: Only $\sim$0.1\% of all objects that
undergo close encounters with a migratory Neptune
have their perihelia raised to avoid
further close encounters over the age of the solar system ({\it Gomes},
2003ab).
Based on this mechanism alone, a disk weighing $\sim$$50 M_{\oplus}$ prior
to migration would have $\sim$$0.05 M_{\oplus}$ deposited into the Scattered
and hot Classical
belts for long-term storage.
But only $\sim$1--2\% of this mass would be in bodies
having sizes $R \gtrsim 100 \km$ ({\it Kenyon and Luu}, 1998, 1999; {\it
Kenyon}, 2002; \S3.1).
Therefore this scenario predicts that
Scattered and hot Classical KBOs having $R \gtrsim 100 \km$ would weigh, in
total,
$\sim$$10^{-3} M_{\oplus}$---about 50--150 times
below what is observed (\S 2.2).
This discrepancy is missed by analyses which neglect consideration of the KBO
size distribution.
A secondary concern is that current numerical simulations of this mechanism
account for the gravitational effects of disk particles
on planets but not on other disk particles. Proper calculation of the
locations
of secular resonances requires, however, a full accounting of
the mass distribution.

Given the low efficiency of the mechanism,
we submit that the high-$q$ orbits of hot Classical and Scattered KBOs
did not arise from Neptune's migration.
Instead, these orbits may have been generated by
Neptune-mass oligarchs whose trajectories passed through the Kuiper belt.
While a numerical simulation is necessary to test
this hypothesis, our order-of-magnitude estimates 
(\S\S 3.3--3.5) for the degree to which oligarchs
stir the belt by simple close encounters are encouraging.
No simulation has yet been performed in which the Kuiper belt is directly
perturbed by a mass as large as
Neptune's for a time as long as $t_{\rm df,cool} \sim 20 \Myr$.
Differences in physical properties between Classical and Scattered/Resonant
KBOs
might still be explained along the same lines as described above:
Scattered/Resonant KBOs were displaced by large distances from
their coagulation zones and so might be expected to exhibit
a large dispersion in color and size, while Classical
KBOs were not so displaced.
Even if all KBOs having $R \gtrsim 100 \km$ were heated
to large $e$ or $i$ by planetary oligarchs, the cold Classical
disk might have re-generated itself in a second wave
of coagulation from a collisional disk of small bodies.

\bigskip
\centerline{3.6.4. Problems Regarding Migration}
\bigskip

The analyses of migration cited above share a common shortcoming: They assume
that planetesimals are collisionless. But coagulation
studies (\S 3.1) indicate that much of the primordial mass remains locked
in small bodies for which collision times threaten to be shorter
than the duration of planetary migration.
By (\ref{tcol}), planetesimals having sizes $\ll 1 \km$ in a minimum-mass disk
have collision times $\ll 20 \Myr$. How Neptune's migration
unfolds when most of the disk comprises highly collisional bodies has not
been well explored. Neptune may open a gap in the disk
(in the same way that moons open gaps in collisional
planetary rings) and the planet's migration may be tied
to how the disk spreads by collisional diffusion ({\it Goldreich et al.},
2004b).

How does the Classical belt shed 99\% of its primordial mass?
Situated at 40--47 AU, it may be too distant for Neptune
to scour directly. Perhaps the small bodies of the Classical belt
are first transported inwards, either by gas drag or collisional diffusion,
and subsequently scoured. Clean-up and migration are intertwined,
but the processes are often not discussed together
(but see {\it Gomes et al.}, 2004).

Are there alternatives to migration for the capture of
Resonant KBOs?
Perhaps Resonant KBOs are captured as Neptune's orbit cools
by dynamical friction (\S 3.5). Before capture,
many belt members would already be stirred to large $e$ and $i$,
not only by unstable oligarchs (\S3.4),
but also by Neptune while it cools.
Cooling accelerates as it proceeds (equation [\ref{dfcool}]). 
A rapid change in the planet's semi-major axis towards the end of cooling
might
trap KBOs into resonance by serendipity. Just after Neptune's
semi-major axis changes, objects having
orbital elements (including longitudes) suitable for libration
would be trapped.
This speculative ``freeze-in'' mechanism
might be too inefficient, since
it requires that the fraction of phase-space volume occupied by resonances
equal
the fraction of KBOs that are Resonant. Taken at face value, observations
suggest the
latter fraction is not much smaller than order unity (\S2.1).

\bigskip
\centerline{\textbf{3.7. Stellar Encounters}}
\bigskip

A passing star may have emplaced Sedna onto its high-perihelion orbit.
For the last $t \sim 4 \Gyr$, solar-mass stars in the Solar neighborhood
have had an average density $n_{\ast} \sim 0.04$ stars pc$^{-3}$
and a velocity dispersion $\langle v_{\ast}^2 \rangle^{1/2} \sim 30 \km
\s^{-1}$.
If we assume that the Sun once resided within a ``typical'' open cluster,
then $n_{\ast} \sim 4$ stars pc$^{-3}$
and $\langle v_{\ast}^2 \rangle^{1/2} \sim 1 \km \s^{-1}$
over $t \sim 200 \Myr$.
Over $t \gtrsim 200 \Myr$, open clusters dissolve by
encounters with molecular clouds ({\it Binney and Tremaine}, 1987).
The number of stars that fly by the Sun within a distance $q_{\ast}$
large enough that gravitational focussing is negligible
($q_{\ast} \gtrsim GM_{\odot}/\langle v_{\ast}^2 \rangle \sim 900 \AU$
for $\langle v_{\ast}^2 \rangle^{1/2} \sim 1 \km \s^{-1}$)
increases as $\int^t n_{\ast} \langle v_{\ast}^2 \rangle^{1/2} dt$.
Therefore fly-bys during the current low-density era outnumber
those during the cluster era by a factor of $\sim$6.
Nonetheless, intra-cluster encounters can be more effective at perturbing
KBO trajectories because encounter velocities are 30 $\times$ lower.

{\it Fern\'andez and Brunini} (2000) simulate the formation of the Oort cloud
within an open cluster having parameters similar to those cited above.
They find that passing stars create an ``inner Oort cloud'' of objects having
$35 \lesssim q(\rm{AU}) \lesssim 1000$,
$300 \lesssim a(\rm{AU}) \lesssim 10^4$,
$\langle e \rangle \sim 0.8$,
and $\langle i^2 \rangle^{1/2} \sim 1$.
Sedna may be the first discovered member of this inner Oort cloud
({\it Brown et al.}, 2004).
Such objects coagulate in the vicinity of the giant planets and
are scattered first by them. Since a scattering event
changes velocities more effectively than it does positions,
objects' perihelia remain at heliocentric distances of $\sim$5--30 AU
while aphelia diffuse outward. Aphelia grow so distant
that objects are scattered next by cluster stars.
These stars raise objects' perihelia beyond the reach of the giant planets.

We confirm the ability of cluster stars to raise the perihelion of Sedna
with an order-of-magnitude calculation.
During the open cluster phase, the number of stars
that pass within distance $q_{\ast}$ of the Sun is

\begin{equation}
N_{\ast} \sim 1 \left( \frac{q_{\ast}}{4000 \AU} \right)^2 \left(
\frac{n_{\ast}}{4 \pc^{-3}} \, \frac{\langle v_{\ast}^2 \rangle^{1/2}}{1 \km
\s^{-1}} \, \frac{t}{200 \Myr} \right) \,.
\end{equation}

\noindent A star of mass $M_{\ast}$ having perihelion distance $q_{\ast}$ much
greater
than a planetesimal's aphelion distance ($Q \approx 2a$)
perturbs that object's specific angular momentum by

\begin{equation}
\delta h = \pm C \frac{GM_{\ast}}{\langle v_{\ast}^2 \rangle^{1/2}} \left(
\frac{a}{q_{\ast}} \right)^2 \,,
\label{yabu}
\end{equation}

\noindent where the numerical coefficient $C$ depends
on the encounter geometry ({\it Yabushita}, 1972). We can derive the form
of (\ref{yabu}) by noting that $\delta h \sim Q \delta v$, where
$\delta v$ is the perturbation to the object's velocity relative to the Sun.
We write $\delta v$ as the tidal acceleration $GM_{\ast} Q / q_{\ast}^3$
induced by the star, 
multiplied by the duration $q_{\ast}/\langle v_{\ast}^2 \rangle^{1/2}$ of the
encounter,
to arrive at (\ref{yabu}). For highly eccentric orbits $\delta q = h \delta h /
(GM_{\odot})$, whence

\begin{equation}
\frac{\delta q}{q} \sim \pm C \frac{M_{\ast}}{M_{\odot}} \left(
\frac{a}{q_{\ast}} \right)^2 \left( \frac{2GM_{\odot}}{q \langle v_{\ast}^2
\rangle} \right)^{1/2} \,.
\end{equation}

\noindent For $M_{\ast} = M_{\odot}$, $q_{\ast} = 4000 \AU$, $C \approx 6$ (see
equation [3.17]
of {\it Yabushita} [1972]), $\langle v_{\ast}^2 \rangle^{1/2} = 1 \km
\s^{-1}$,
and pre-encounter values of $q = 35 \AU$ and $a = 600 \AU$, 
$\delta q / q \sim \pm 1$. Thus, Sedna's perihelion could have doubled
to near its current value, $q \approx 76 \AU$, by a single slow-moving cluster
star.
Multiple encounters at larger $q_{\ast}$ cause $q$ to random walk
and change its value less effectively:
$\langle (\delta q)^2 \rangle^{1/2} \propto (N_{\ast})^{1/2} q_{\ast}^{-2}
\propto q_{\ast}^{-1}$.

Had we performed this calculation for parameters appropriate to the
present-day
stellar environment, we would have found $\delta q / q \approx \pm 0.2$.
The reduction in efficacy is due to the larger $\langle v_{\ast}^2 \rangle$
today.

The cluster properties cited above are averaged over a 
half-light radius of 2 pc ({\it Binney and Tremaine}, 1987).
For comparison, the Hyades cluster has
$4 \times$ lower $n_{\ast}$, $3\times$ lower $\langle v_{\ast}^2
\rangle^{1/2}$,
and $6 \times$ longer lifetime $t$ ({\it Binney and Merrifield}, 1998; {\it
Perryman et al.}, 1998);
the Hyades therefore generates $2 \times$ fewer encounters than does our
canonical cluster.
Younger clusters like the Orion Trapezium maintain $15 \times$ higher
$n_{\ast}$
and similar $\langle v_{\ast}^2 \rangle^{1/2}$ over $200 \times$ shorter $t$
({\it Hillenbrand and Hartmann}, 1998), and therefore yield even fewer
encounters.
Scenarios that invoke stellar encounters for which $q_{\ast} \ll 1000 \AU$
to explain such features as the edge of the Classical belt
require that the Sun have resided in a cluster
having atypical properties, i.e., dissimilar from those of the
Orion Trapezium, the Hyades, and all open clusters documented by {\it Binney
and Merrifield}
(1998). That parent bodies in extra-solar debris disks also do not extend
beyond
$\sim$40--100 AU (\S3.1.2)
argues against explanations that rely on unusually dense environments.

\bigskip
\centerline{{\bf 3.8. Coagulation of Neptune Trojans}}
\bigskip

Planetesimal collisions that occur
near Neptune's Lagrange points insert
debris into 1:1 resonance. This debris
can coagulate into larger
bodies. The problem of accretion in the Trojan resonance is akin
to the standard problem of planet formation,
transplanted from a star-centered disk to a disk
centered on the Lagrange point. As with
other kinds of transplant operations, there
are complications: Additional timescales not present
in the standard problem, such as the
libration period $T_{\rm lib}$ about the Lagrange point, require juggling.
{\it Chiang and Lithwick} (2005, hereafter {\it CL}05) account for these
complications
to conclude that QR-sized Trojans may form as
miniature oligarchs, each dominating
its own tadpole-shaped annulus in the ancient Trojan sub-disk.
Alternative formation scenarios for Trojans such
as pull-down capture and direct collisional emplacement
of QR-sized objects into resonance are considered by {\it CL}05
and deemed unlikely. Also, the mechanism proposed by {\it Morbidelli et
al.}~(2005)
to capture Jupiter Trojans cannot be applied to Neptune Trojans
since Uranus and Neptune today lie inside their 1:2 MMR and
therefore could not have divergently migrated across it 
({\it Morbidelli}, personal communication). We focus on {\it in situ}
accretion, but acknowledge that a collisionless
capture scenario might still be feasible and even favored
by late-breaking data; see the end of this sub-section.

In the theory of oligarchic planet formation (e.g., {\it G}04),
each annulus is of order $5R_{\rm H}$ in radial width;
the number of QR-sized oligarchs that can be fitted
into the tadpole libration region is

\begin{equation}
N_{\rm Trojan} \sim \frac{(8M_{\rm N}/3M_{\odot})^{1/2}a_{\rm N}}{5R_{\rm H}}
\sim 20 \,,
\label{attractive}
\end{equation}

\ni attractively close to the number of QR-sized
Neptune Trojans inferred to exist today (\S 2.2.3).
The numerator in (\ref{attractive}) equals the maximum
width of the 1:1 MMR, $a_{\rm N} \approx 30 \AU$
is Neptune's current semi-major axis, $R_{\rm H} = R/\alpha$
is the Trojan's Hill radius, and $R \approx 90 \km$ is the radius of QR.

The input parameters of the coagulation model are the surface density $\sigma$
and sizes $s$ of small bodies in 1:1 resonance.
Big bodies grow by consuming small bodies, but growth is limited
because small bodies diffuse out of resonance by colliding with other small
bodies.
The time for a small body to random walk out of the Trojan sub-disk is

\begin{equation}
t_{\rm esc} \sim \frac{\rho s}{\sigma \Omega} \left[ \frac{(M_{\rm
N}/M_{\odot})^{1/2}a_{\rm N}}{u/\Omega} \right]^2 \,.
\label{tesc}
\end{equation}

\ni The term in square brackets follows from noting that a small body
shifts its orbital guiding center by of order its epicyclic amplitude
$\sim$$\pm u/\Omega$
every time it collides with another small body in an optically thin disk.
To escape resonance, the small body must random walk the maximum libration
width.
We equate $t_{\rm esc}$ to the growth time of a big body $t_{\rm acc}$
(equation [\ref{rrdot}])
to solve for the maximum size to which a large body coagulates:

\begin{equation}
R = R_{\rm final} \sim 100 \left( \frac{2}{u/v_{\rm H}} \right)^{4/3} \left(
\frac{s}{20 \cm} \right)^{1/3} \km \,.
\label{rfinal}
\end{equation}

\ni 
Our normalization
of $u/v_{\rm H} \approx 2$ is derived from $s \sim 20 \cm$
and $\sigma \sim 4 \times 10^{-4} \gm \cm^{-2} \sim$
10 times the surface density inferred in QR-sized
objects today; we derive $u/v_{\rm H}$
by balancing gravitational stirring by big bodies with
damping by inelastic collisions between small bodies ({\it CL}05).
For these parameter values,
$t_{\rm esc} \sim t_{\rm acc} \sim 1 \times 10^9 \yr$.
Unlike Neptune-sized oligarchs that may have
been ejected out of the solar system (\S3.4),
all $\sim$10--30 Trojan oligarchs in a single cloud should be present and
eventually
accounted for. 

As speculated by {\it CL}05, orbital inclinations of Trojans
with respect to Neptune's orbit plane might be small;
perhaps $\langle i^2 \rangle^{1/2} \lesssim 10^{\circ}$.
A thin disk of Neptune Trojans would contrast with the thick disks
occupied by Jupiter Trojans, main belt
asteroids, and non-Classical KBOs, and would reflect
a collisional, dissipative birth environment.
Three other Neptune Trojans have since been
announced after the discovery of QR, having
inclinations of 1.4$^{\circ}$,
25.1$^{\circ}$, and 5.3$^{\circ}$ ({\it Sheppard and Trujillo}, 2006).
If a large fraction of Neptune Trojans have high $i$, we might look to
the $\nu_{18}$ secular resonance, unmodelled by {\it CL}05,
to amplify inclinations. See also {\it Tsiganis et al.}~(2005)
who find that Neptune Trojans can be captured collisionlessly;
the capture process is related to ``freeze-in'' as described in \S3.6.4.

\bigskip
\centerline{{\bf 3.9. Collisional Comminution}}
\bigskip

Over the last few billion years, sufficiently
small and numerous bodies in the Kuiper belt suffer collisional attrition.
As interpreted by {\it Pan and Sari} (2005, hereafter {\it PS}05), the break in
the size
distribution of KBOs at $R \approx 50 \km$ as measured by
{\it Bernstein et al.}~(2004; \S2.2.1) divides the collisional spectrum at
small $R$
from the primordial coagulation spectrum at large $R$.
For the remainder of this subsection, we do not distinguish between
the various dynamical classes but instead analyze
all KBOs together as a single group.
At $R > R_{\rm break}$,
the size spectrum $dN/dR \propto R^{-\tilde{q}_0}$, where
$dN$ is the number of objects per unit face-on
area of the belt having sizes between $R$ and $R+dR$ (the differential
surface number density). The slope $\tilde{q}_0 \sim 5$ (see \S2.2.1 for more
precise values)
presumably represents the unadulterated outcome of coagulation.
Bodies at this large-$R$ end of the spectrum are insufficiently
numerous to collide amongst themselves and undergo attrition.
At $R < R_{\rm break}$, $dN/dR \propto R^{-\tilde{q}}$,
where $\tilde{q}$ derives from a quasi-steady collisional
cascade ({\it Dohnanyi}, 1969; {\it PS}05).
By definition of $R_{\rm break}$, the time for a body
of radius $R_{\rm break}$ to be catastrophically dispersed
equals the time elapsed:

\begin{equation}
\frac{1}{N_{\rm proj} \times \pi R_{\rm break}^2 \times \Omega} \sim t \,,
\end{equation}

\ni where $\pi R_{\rm break}^2$ is the collision
cross-section and $N_{\rm proj}$ is the surface number density
of projectiles that are just large enough to disperse $R_{\rm break}$-sized
targets (catastrophic dispersal
implies that the mass of the largest post-collision fragment
is no greater than half the mass of the original target and
that collision fragments disperse without gravitational reassembly).
This expression is valid for the same assumptions
underlying equation (\ref{tcol}), i.e., for today's dynamically hot belt.

We proceed to estimate
$R_{\rm break}$ given the parameters of the present-day Kuiper belt.
For $R > R_{\rm break}$, $N = N_0 (R/R_0)^{1-\tilde{q}_0}$, where
$N$ is the surface number density of objects having sizes
between $R$ and $2R$. We estimate
that for fiducial radius $R_0 = 100\km$, $N_0 \approx 20 \AU^{-2}$
at $a \approx 43 \AU$.
The minimum radius $R_{\rm proj}$ of the projectile that can catastrophically
disperse
a target of radius $R_{\rm break}$ is given by

\begin{equation}
\frac{1}{2} R_{\rm proj}^3 v_{\rm rel}^2 = R_{\rm break}^3 Q^{\ast}
\end{equation}

\ni where

\begin{equation}
Q^{\ast} = Q^{\ast}_0 \left( \frac{R}{R_0} \right)^y
\end{equation}

\ni is the collisional specific energy ({\it Greenberg et al.}, 1978; {\it
Fujiwara et al.}, 1989)
and $v_{\rm rel}$ is the relative collision velocity. Since for $R < R_{\rm
break}$
as much mass is ground into every logarithmic interval in $R$ as is ground out
(e.g., {\it PS}05),

\begin{equation}
\tilde{q} = \frac{21+y}{6+y} \,.
\end{equation}

We assume (and can check afterwards) that $R_{\rm proj} < R_{\rm break} < R_0$
to write

\begin{equation}
N_{\rm proj} = N_0 \left( \frac{R_{\rm break}}{R_0} \right)^{1-\tilde{q}_0}
\left( \frac{R_{\rm proj}}{R_{\rm break}} \right)^{1-\tilde{q}} \,.
\end{equation}

\ni Combining the above relations yields

\begin{eqnarray}
\frac{R_{\rm break}}{R_0} \sim \left( \pi N_0 R_0^2 \Omega t \right)^{z_1}
\left( \frac{v_{\rm rel}^2}{2Q^{\ast}_0} \right)^{z_2}
\,,
\label{whopper}
\end{eqnarray}

\ni where $z_1 = (6+y)/[5y + (6+y)(\tilde{q}_0-3)]$ and
$z_2 = 5/[5y + (6+y)(\tilde{q}_0-3)]$.
For targets held together by self-gravity, $Q^{\ast} \approx 3 v_{\rm esc}^2 /
10$
and $y = 2$. If we insert these values into (\ref{whopper}), together
with $v_{\rm rel} = 1 \km \s^{-1}$, $\tilde{q}_0 = 5$, $\Omega = 2\pi/(300
\yr)$,
and $t = 3 \times 10^9 \yr$, we find that
$R_{\rm break} \approx 0.4 R_0 \approx 40 \km$, in good
agreement with the observed break in the luminosity function (Fig.~5; {\it
PS}05).
The small-$R$ end of the KBO size spectrum as observed today
reflects the catastrophic comminution of bodies that derive
their strength from self-gravity (``rubble piles'').
Furthermore, the Kuiper belt has been dynamically hot
for the last few billion years ({\it PS}05).

\bigskip
\centerline{\textbf{4. DIRECTIONS FOR FUTURE WORK}}
\bigskip

1. {\it Collisional vs.~Collisionless:} Most explorations
of planetary migration and of how the Kuiper belt was stirred utilize
collisionless gravitational simulations. But the overwhelming bulk
of the primordial
mass may have resided in small, collisional bodies.
Simultaneously accounting for collisions and gravity
might revolutionize our understanding of the
clean-up (a.k.a.~missing-mass) problem.
Insights from the study of planetary rings will be helpful.

2. {\it Classical KBO Colors vs.~Heliocentric Distance:}
Do Classical KBOs exhibit a trend in color from neutral to red with increasing
heliocentric distance $d$?
The two neutral Classicals at $d \approx 38 \AU$, contrasted with the
predominantly
red Classicals at $d \approx 42 \AU$, suggest the answer is yes (Figs.~3 and
4).
Confirmation would support ideas that Classicals coagulated
{\it in situ}, and that neutrally colored Resonant/Scattered
KBOs coagulated from small $d$ and were transported outwards.
We must also ask why trends in color with birth
distance $d$ would exist in the first place.

3. {\it Formation of the Scattered Belt by Neptune-Mass Oligarchs:}
We argue that Neptune's migration and the concomitant sweeping of secular
resonances do not populate the Scattered and hot Classical belts with enough
objects
to explain observations. When account is made of the
primordial size distribution of planetesimals---a distribution that should be
preserved today at large sizes (\S1; \S3.9)---the
expected population of Scattered / hot Classical objects having sizes above 100
km
is less than that observed by a factor of 50--150.
We propose instead that planetesimals were deflected onto Scattered / hot
Classical orbits
by simple close encounters with
marauding Neptune-mass oligarchs that have since been ejected
from the solar system, and by Neptune while its orbit circularized by dynamical
friction.
These contentions are supported by order-of-magnitude estimates
but require numerical simulations to verify.

4. {\it Kuiper Cliff:} Why do planetesimal disks have sharp outer edges?

5. {\it Binaries}: Kuiper belt binaries might prove the most informative
witnesses we have to the history of trans-Neptunian space.
They hearken back to a primordially dense and cold disk in which collisions
and multiple-body encounters were orders of magnitude more frequent than they
are
today. Binary orbit properties must also reflect how the Kuiper belt was
stirred as a whole.
How binary inclinations,
eccentricities, and component mass ratios are distributed,
and how/why the incidence of binarity correlates with dynamical class
are open issues for observer and theorist alike.

\textbf{Acknowledgments.} This work was supported by
the Alfred P.~Sloan Foundation, NSF, and NASA. We acknowledge helpful
exchanges with G.~Bernstein, P.~Goldreich, R.~Gomes, D.~Jewitt, S.~Kenyon,
A.~Morbidelli, D.~Nesvorn\'y,
R.~Sari, L.~Strubbe, A.~Youdin, and an anonymous referee.
We are grateful to the Deep Ecliptic Survey (DES)
team for their unstinting support.

\bigskip

\centerline\textbf{REFERENCES}
\bigskip
\parskip=0pt
{\small
\baselineskip=11pt
\refs Bernstein G.~M., Trilling D.~E., Allen R.~L., Brown M.~E., Holman M., and
Malhotra R.~(2004) {\it Astron.~J., 128}, 1364--1390 ({\it B}04).
\refs Boehnhardt H., Tozzi G.~P., Birkle K., Hainaut O., Sekiguchi T.,
   et al.~(2001)
  {\it Astron.~Astrophys., 378}, 653--667.
\refs Binney J. and Tremaine S. (1987) {\it Galactic Dynamics}, pp.~26,
440--443. Princeton University, Princeton.
\refs Binney J. and Merrifield M. (1998) {\it Galactic Astronomy},
pp.~377--386. Princeton University, Princeton.
\refs Brown M.~E., Trujillo C., and Rabinowitz D. (2004) {\it Astrophys.~J.,
617}, 645--649.
\refs Brown M.~E., Trujillo C.~A., and Rabinowitz D.~L. (2005a), {\it
Astrophys.~J., 635}, L97--100.
\refs Brown M.~E., van Dam M.~A., Bouchez A.~H., Le Mignant D., Campbell R.~D.,
et al.~(2005b)
{\it Astrophys.~J., 632}, L45--48. 
\refs Buie M.~W., Grundy W.~M., Young E.~F., Young, L.~A., and Stern
S.~A.~(2006) {\it Astron.~J., 132}, 290--298.
\refs Canup R.~M. (2005) {\it Science, 307}, 546--550.
\refs Chiang E.~I. and Jordan A.~B. (2002) {\it Astron.~J., 124}, 3430.
\refs Chiang E.~I., Jordan A.~B., Millis R.~L., Buie M.~W., Wasserman L.~H., et
al.~(2003)
{\it Astron.~J., 126}, 430--443 ({\it C}03).
\refs Chiang E.~I. and Lithwick Y.~(2005) {\it Astrophys.~J., 628}, 520--532
({\it CL}05).
\refs Christy J.~W. and Harrington R.~S. (1978) {\it Astron.~J., 83},
1005--1008.
\refs Dermott S.~F., Malhotra R., and Murray C.~D. (1988) {\it Icarus, 76},
295--334.
\refs Dohnanyi J.~W. (1969) {\it J.~Geophys.~Res., 74}, 2531--2554.
\refs Duncan M.~J., Levison H.~F., and Budd S.~M. (1995) {\it Astron.~J., 110},
3073--3081 ({\it D}95).
\refs Elliot J.~L., Kern S.~D., Clancy K.~B., Gulbis A.~A.~S., Millis R.~L., et
al.~(2005)
{\it Astron.~J., 129}, 1117--1162 ({\it E}05).
\refs Fern\'andez J.~A. and Ip W.-H. (1984) {\it Icarus, 58}, 109--120.
\refs Fern\'andez J.~A. and Brunini A. (2000) {\it Icarus, 145}, 580--590.
\refs Fujiwara A., Cerroni P., Davis D., Ryan E., and di Martino M. (1989) In
{\it Asteroids II} (R.~P.~Binzel et al., eds.), pp.~240--265. Univ.~of Arizona,
Tucson.
\refs Funato Y., Makino J., Hut P., Kokubo E., and Kinoshita D. (2004) {\it
Nature, 427}, 518--520.
\refs Garaud P. and Lin D.~N.~C. (2004) {\it Astrophys.~J., 608}, 1050--1075.
\refs Gladman B., Holman M., Grav T., Kavelaars J., Nicholson P., et
al.~(2002)
{\it Icarus, 157}, 269--279.
\refs Goldreich P. (1965) {\it Mon.~Not.~R.~Astron.~Soc., 130}, 159--181.
\refs Goldreich P., Lithwick Y., and Sari R. (2002) {\it Nature, 420}, 643--646
({\it G}02).
\refs Goldreich P., Lithwick Y., and Sari R. (2004a) {\it
Ann.~Rev.~Astron.~Astrophys., 42}, 549--601 ({\it G}04).
\refs Goldreich P., Lithwick Y., and Sari R. (2004b) {\it Astrophys.~J., 614},
497--507.
\refs Gomes R.~S. (2003a) {\it Icarus, 161}, 404--418.
\refs Gomes R.~S. (2003b) {\it Earth Moon Planets, 92}, 29--42.
\refs Gomes R.~S., Morbidelli A., and Levison H.~F. (2004) {\it Icarus, 170},
492--507.
\refs Gomes R.~S., Gallardo T., Fern\'andez J.~A., and Brunini A. (2005) {\it
Cel.~Mech.~Dyn.~Astron., 91}, 109--129.
\refs G\'omez G.~C. and Ostriker E.~C. (2005) {\it Astrophys.~J., 630},
1093--1106.
\refs Greenberg R., Hartmann W.~K., Chapman C.~R., and Wacker J.~F. (1978) {\it
Icarus, 35}, 1--26.
\refs Greenberg R., Bottke W.~F., Carusi A., and Valsecchi G.~B. (1991) {\it
Icarus, 94}, 98--111.
\refs Hahn J.~M. and Malhotra R. (1999) {\it Astron.~J., 117}, 3041--3053.
\refs Hahn J.~M. and Malhotra R. (2005) {\it Astron.~J., 130}, 2392--2414.
\refs Hillenbrand L.~A. and Hartmann L.~W. (1998) {\it Astrophys.~J., 492},
540--553.
\refs Holman M.~J. and Wisdom J. (1993) {\it Astron.~J., 105}, 1987--1999 ({\it
HW}93).
\refs Ida S. and Makino J. (1993) {\it Icarus, 106}, 210--217.
\refs Jewitt D. and Luu J. (1993) {\it Nature, 362}, 730--732.
\refs Jewitt D. and Luu J. (2000) In {\it Protostars and Planets IV}
(V.~Mannings et al., eds.), pp.~1201--1229. Univ.~of Arizona, Tucson.
\refs Jewitt D., Luu J., and Trujillo C. (1998) {\it Astron.~J., 115},
2125--2135.
\refs Kalas P. and Jewitt D. (1995) {\it Astron.~J., 110}, 794--804.
\refs Kenyon S.~J. and Bromley B.~C. (2006) {\it Astrophys.~J., 131},
1837--1850.
\refs Kenyon S.~J. (2002) {\it Publ.~Astron.~Soc.~Pac., 793}, 265--283.
\refs Kenyon S.~J. and Luu J.~X. (1998) {\it Astron.~J., 115}, 2136--2160.
\refs Kenyon S.~J. and Luu J.~X. (1999) {\it Astron.~J., 118}, 1101--1119.
\refs Kne\v{z}evi\'{c} Z., Milani A., Farinella P., Froeschle Ch., and
Froeschle Cl. (1991) {\it Icarus, 93}, 315--330.
\refs Krist J.~E., Ardila D.~R., Golimowski D.~A., Clampin M., Ford H.~C., et
al.~(2005) {\it Astron.~J., 129}, 1008--1017.
\refs Levison H.~F. and Morbidelli A. (2003) {\it Nature, 426}, 419--421.
\refs Malhotra R. (1993) {\it Nature, 365}, 819--821.
\refs Malhotra R. (1995) {\it Astron.~J., 110}, 420--429.
\refs Matsuyama I., Johnston D., and Hartmann L. (2003) {\it Astrophys.~J.,
582}, 893--904.
\refs Millis R.~L., Buie M.~W., Wasserman L.~H., Elliot J.~L., Kern S.~D., and
Wagner R.~M. (2002) {\it Astron.~J., 123}, 2083--2109.
\refs Morbidelli A., Jacob C., and Petit J.-M. (2002) {\it Icarus, 157},
241--248.
\refs Morbidelli A., Levison H.~F., Tsiganis K., and Gomes R. (2005) {\it
Nature, 435}, 462--465.
\refs Murray C.~D. and Dermott S.~F. (1999) {\it Solar System Dynamics},
pp.~63--128, 225--406. Cambridge University, Cambridge.
\refs Murray-Clay R.~A. and Chiang E.~I. (2005) {\it Astrophys.~J., 619},
623--638 ({\it MC}05).
\refs Murray-Clay R.~A. and Chiang E.~I. (2006) {\it Astrophys.~J.}, in press
(astro-ph/0607203) ({\it MC}06).
\refs Nesvorn\'y D. and Dones L. (2002) {\it Icarus, 160}, 271--288.
\refs Noll K. (2003) {\it Earth Moon Planets, 92}, 395--407.
\refs Pan M. and Sari R. (2004) {\it Astron.~J., 128}, 1418--1429.
\refs Pan M. and Sari R. (2005) {\it Icarus, 173}, 342--348 ({\it PS}05).
\refs Peale S.~J. (1986) In {\it Satellites}, (J.~A.~Burns and M.~S.~Matthews,
eds.), pp.~159--223. Univ.~of Arizona, Tucson.
\refs Peixinho N., Boehnhardt H., Belskaya I., Doressoundiram A., Barucci
M.~A., and Delsanti A. (2004) {\it Icarus, 170}, 153--166.
\refs Perryman M.~A.~C., Brown A.~G.~A., Lebreton Y., Gomez A., Turon C., et
al.~(1998)
{\it Astron.~Astrophys., 331}, 81--120.
\refs Press W.~H., Teukolsky S.~A., Vetterling W.~T., and Flannery B.~P. (1992)
{\it Numerical Recipes in C: The Art of Scientific Computing}, pp.~609--639.
Cambridge University, Cambridge.
\refs Sheppard S.~S. and Jewitt D. (2004) {\it Astron.~J., 127}, 3023--3033.
\refs Sheppard S.~S. and Trujillo C.~A. (2006) {\it Science, 313}, 511--514.
\refs Stephens D.~C. and Noll K.~S. (2006) {\it Astron.~J., 131}, 1142--1148.
\refs Stern S.~A., Weaver H.~A., Steffl A.~J., Mutchler M.~J., Merline W.~J.,
et al.~(2006) {\it Nature, 439}, 946--948.
\refs Strubbe L.~E. and Chiang E.~I. (2006) {\it Astrophys.~J., 648},
652--665.
\refs Thommes E.~W., Duncan M.~J., and Levison H.~F. (1999) {\it Nature, 402},
635--638.
\refs Thommes E.~W., Duncan M.~J., and Levison H.~F. (2002) {\it Astron.~J.,
123}, 2862--2883.
\refs Tiscareno M.~S. and Malhotra R. (2003) {\it Astron.~J., 126},
3122--3131.
\refs Trujillo C.~A. and Brown M.~E. (2001) {\it Astrophys.~J., 554}, L95--98.
\refs Trujillo C.~A. and Brown M.~E. (2002) {\it Astrophys.~J., 566},
L125--128.
\refs Tsiganis K., Gomes R., Morbidelli A., and Levison H.~F. (2005) {\it
Nature, 435}, 459--461.
\refs Veillet C., Parker J.~W., Griffin I., Marsden B., Doressoudiram A., et
al.~(2002)
{\it Nature, 416}, 711--713.
\refs Weaver H.~A., Stern S.~A., Mutchler M.~J., Steffl A.~J., Buie M.~W., et
al.~(2006) {\it Nature, 439}, 943--945.
\refs Weidenschilling S.~J. (2002) {\it Icarus, 160}, 212--215.
\refs Yabushita S. (1972) {\it Astron.~Astrophys., 16}, 395--403.
\refs Youdin A.~N. and Shu F.~H. (2002) {\it Astrophys.~J., 580}, 494--505.
\refs Youdin A.~N. and Chiang E.~I. (2004) {\it Astrophys.~J., 601},
1109--1119.
\refs Youdin A.~N. and Goodman J. (2005) {\it Astrophys.~J., 620}, 459--469.



\end{document}